\documentclass[fleqn]{2026SCGE}
\usepackage{bm}
\usepackage{algorithm}
\usepackage{hyperref}
\usepackage{amsmath}
\usepackage{cuted}
\usepackage[normalem]{ulem}
\usepackage{soul,xcolor}

\begin{document}
\ensubject{subject}

\ArticleType{Article}
\Year{2026}
\Month{January}
\Vol{69}
\No{1}
\DOI{??}
\ArtNo{000000}
\ReceiveDate{2026}
\AcceptDate{2026}

\title{Probing an Intermediate-Mass Black Hole Companion of Sagittarius~A* \\with Pulsar Timing}

\author[1, 2]{Zexin Hu}{}%
\author[2, 3, \thanks{Corresponding author (email: {\color{blue}lshao@pku.edu.cn})}]{Lijing Shao}{}

\AuthorMark{Zexin Hu \& Lijing Shao}

\AuthorCitation{Zexin Hu \& Lijing Shao}

\address[1]{Department of Astronomy, School of Physics, Peking University, Beijing 100871, China}
\address[2]{Kavli Institute for Astronomy and Astrophysics, Peking University, Beijing 100871, China}
\address[3]{National Astronomical Observatories, Chinese Academy of Sciences, Beijing 100101, China}


\abstract{

An intermediate-mass black hole (IMBH) hidden in our Galactic Center (GC) may explain the puzzling observations of the stellar distribution around Sagittarius~A* (Sgr~A*), the supermassive black hole (SMBH) in the GC. Future observations with the next-generation radio telescopes, such as the SKA, are promising to discover pulsars orbiting around Sgr~A*, and thus provide the possibility of constraining the hidden IMBH with pulsar timing. We study the detectability of a third-body, the IMBH, in the pulsar-SMBH system based on radio timing observation. We find that the pulsar-SMBH system is very sensitive to such a third-body perturbation and can be used to put stringent constraints on the existence of the IMBH. Even under strong perturbations caused by the complex GC astrophysical environments, timing observation will still complement the existing observational constraints.

}


\maketitle


\begin{multicols}{2}
\section{Introduction}\label{section1}

Observed stellar distributions around the supermassive black hole (SMBH), Sagittarius~A* (Sgr~A*), in our Galactic Center (GC) are not straightly explained by the normal star formation models, especially for the population of the so-called S-stars, which seem to be too young to present at their current position~\cite{Ghez:2003rt}, while there is lack of more massive bodies segregated closer to the GC~\cite{Alexander:2008tq}. The coexistence of the clockwise-rotating stellar-disk stars and the off-disk stars further requires a complex model to explain them simultaneously. One possible way to explain the age and location of the S-stars requires an intermediate-mass black hole (IMBH) in the GC~\cite{Hansen:2003yb}, which can drag the S-stars inwards and explain their large velocity dispersion~\cite{Yu:2007jz,Zheng:2026}.  Though not all scenarios need an IMBH~\cite{Chen:2014dya}, it is still natural to ask whether Sgr~A* has a hidden companion due to the hierarchical nature of the widely accepted galaxy formation paradigm~\cite{Hopkins:2005fb}. Remnant IMBHs in the GC can be the result of minor merger with low-mass dwarf galaxies or globular clusters~\cite{Rashkov:2013uua}. Current observations, as well as theoretical arguments, have not excluded this possibility yet~\cite{Naoz:2019sjx, Zhang:2023ekp, Will:2023nlt, GRAVITY:2023met}.

One of the strongest constraints on the existence of a hidden IMBH in the GC comes from the 23 years of orbit tracing of the star S0-2~\cite{Will:2023nlt, GRAVITY:2023met}. The data excluded an IMBH with mass larger than $10^3\, M_\odot$ and an orbital semi-major axis similar to that of the S0-2 ($\sim 1020\, {\rm AU}$), though slightly different conclusions were obtained by two groups~\cite{Will:2023nlt, GRAVITY:2023met}. Additional constraints on the companion of Sgr~A* result from the proper motion measurements of Sgr~A*~\cite{Hansen:2003yb,Yu:2003hj,Reid:2020}, as well as modeling the stability and distribution of the S-star cluster~\cite{Zhang:2023ekp, Naoz:2019sjx}. It is expected that future gravitational-wave observations with the space-borne detectors or the pulsar timing arrays could put tighter constraints on the presence of an IMBH in the GC~\cite{Strokov:2023kmo, Guo:2024tlg}.

Timing a pulsar orbiting around Sgr~A* could be another powerful probe to detect the existence of a hidden IMBH in the GC. Previous studies have shown that timing an ideal pulsar in the GC could provide precise measurements of the properties of the Sgr~A* and give several unique tests of general relativity (GR) or dark matter models~\cite{Wex:1998wt, Kramer:2004hd, Liu:2011ae, Psaltis:2015uza, Hu:2023ubk, Dong:2022zvh, Hu:2023vsg, DellaMonica:2023ydm, Hu:2024blq, Shao:2025vmb, DellaMonica:2025ent, Yu:2025apk, Hu:2026zcb}. Meanwhile, the high measurement precision in timing observation also allows the detection of the astrophysical environment around the Sgr~A*~\cite{Hu:2023ubk}. The existence of an IMBH in the GC would perturb the motion of the orbiting pulsar and leave detectable imprints in the timing residuals. In fact, any significant mass distributions around Sgr~A* would affect the dynamics of the pulsar-SMBH system. For example, the stellar mass black hole (BH) cusp around the SMBH could cause a strong background noise in the timing data that is crucial for analysis~\cite{Hu:2026aez}. Nevertheless, if an IMBH is located in a proper location, one would expect it to provide the loudest residual signal above the background noise and with characteristic features.

Similar to the S-star cluster~\cite{Naoz:2019sjx}, the long-term dynamical stability of a pulsar-SMBH-IMBH system can also place constraints 
on the allowed properties and orbital configurations of the IMBH. However, in this work, we aim to provide an independent constraint using 
pulsar timing and therefore do not impose additional constraints from the long-term dynamical stability of the system.

Despite theoretical models and observational evidences suggest that there could be a large pulsar population hiding in the GC~\cite{Pfahl:2003tf, Zhang:2014kva, Schoedel2020, SKAOPulsarScienceWorkingGroup:2025syv}, currently there are only seven pulsars been found in the inner $100\,{\rm pc}$ of the GC~\cite{Johnston:2006fx, Deneva:2009mx, Eatough:2013nva, Rea:2013pqa, Wongphechauxsorn:2023qcy}, and none of them are close enough to perform the measurements mentioned above. The lack of observed pulsars in the GC might be explained by the complex interstellar medium environment in that region~\cite{Yao2017, Ocker:2026mta}. Common observations at the low frequencies suffer from strong dispersion and scattering, while previous high-frequency surveys might be limited by the steep spectrum of the pulsar emission, which requires a high instrument sensitivity~\cite{Liu:2021ziv, EventHorizonTelescope:2023atv}. It is expected that future observation with the next-generation radio telescopes, like the SKA, could find a number of pulsars in the GC and even pulsars that are suitable for the proposed measurements mentioned before~\cite{SKAOPulsarScienceWorkingGroup:2025syv}.

The high measurement precision of pulsar timing allows the detection of relativistic effects. Unlike the Keplerian orbit in the Newtonian dynamics, binary motion in full GR does not have a closed analytic solution yet. For timing observation of normal binary pulsar systems~\cite{Kramer:2021jcw, Hu:2023vsq},  it is usually enough to use the first post-Newtonian (PN) approximation to describe the orbital motion, for which an elegant quasi-Keplerian solution is available~\cite{Damour:1985}. However, for the pulsar-SMBH system we consider here, possibly with an additional IMBH to form a three-body system, one can only rely on numerical integration of the orbital motion. A similar approach was adopted for the timing of  PSR~J0337$+$1715 in a triple stellar system~\cite{Shao:2016ubu, Voisin:2020lqi}.

In this paper, we focus on the detectability of an IMBH companion of Sgr~A* with pulsar timing observation. We construct a numerical timing model for the pulsar-SMBH-IMBH system based on the canonical equations of motion derived from the 1\,PN Hamiltonian~\cite{Heinze:2026rho, Einstein:1938yz}. We take into account the spin-orbit interaction caused by the SMBH and the three-body interacting term in the equations of motion~\cite{Will:2013cza}. Based on the timing model, we study the various possible timing residuals caused by the presence of the IMBH, which could reveal the existence of a hidden companion of Sgr~A* if such characteristic signals are observed in real timing observations. Using the Fisher matrix approximation, we forecast the measurability of the IMBH parameters in an ideal case. Considering the complex astrophysical environment around Sgr~A*, we also give an estimation of the IMBH detectability fully based on the timing residual. 

The remaining paper is organized as follows. In Sec.~\ref{sec:timing model}, we construct the numerical timing model for the pulsar-SMBH-IMBH system. In Sec.~\ref{sec:time delays}, we present various timing residuals caused by the IMBH. We forecast the measurement precision of IMBH parameters using the Fisher matrix approximation in Sec.~\ref{sec:PE}, and estimate the IMBH detectability for more realistic cases in Sec.~\ref{sec:detect}. Finally, we conclude in Sec.~\ref{sec:discussion}.


\section{Timing model}\label{sec:timing model}

Data analysis of pulsar timing observation relies on the so-called timing model, 
which relates the proper rotation of the pulsar to the times of arrival (TOAs) of pulsar pulses observed by the 
radio telescopes~\cite{Damour:1986}.  The rotation stability of neutron stars enables high-precision measurements 
in the pulsar timing. To build a timing model for a pulsar, one needs to solve the pulsar's orbital motion and the 
light propagation in the curved spacetime of its companion(s). In Sec.~\ref{subsec:orbital motion}, we discuss 
the motion of the three-body system at the 1\,PN level. The various time delays related to the light propagation 
are introduced in Sec.~\ref{subsec:light propagation}. For the purpose of forecasting the detectability of the 
IMBH, the timing model we constructed here only takes into account various leading-order effects. We also ignore 
the contributions from the Solar system~\cite{Lorimer:2004handbook}, which should be incorporated in real observations.

\subsection{Orbital Motion}\label{subsec:orbital motion}

We focus on an idealized system that contains only three bodies: a SMBH (i.e.,~Sgr\,A*), an IMBH, and a pulsar. We calculate the orbital motion of the pulsar by numerically integrating the PN equations of motion of the system. Here, we consider the canonical equations of motion derived from the 1\,PN Hamiltonian with leading-order spin-orbit interaction, 
\begin{equation}
  H=H_{\rm N}+H_{\rm 1PN}+H_{\rm SO}\,.
\end{equation}
These equations explicitly conserve the total energy and momenta. In the Arnowitt-Deser-Misner (ADM) coordinates and with the spin supplementary condition (SSC) of Pryce~\cite{Pryce:1948pf} and of Newton and Wigner~\cite{Newton:1949cq}, one has~\cite{Heinze:2026rho, Einstein:1938yz, Wex:1995pjg}
\begin{align}
  H_{\rm N}&=\frac{1}{2}\sum_a\frac{p_a^2}{m_a}-\frac{1}{2}\sum_a\sum_{b\neq a}\frac{Gm_a m_b}{r_{ab}}\,,\\
  c^2 H_{\rm 1PN}&=-\frac{1}{8}\sum_a \frac{p_a^4}{m_a^3}-\frac{1}{4}\sum_a\sum_{b\neq a}\frac{Gm_a m_b}{r_{ab}}
  \left[6\frac{p_a^2}{m_a^2} \right.\nonumber\\
  &\left.-7\frac{\bm{p}_a\cdot\bm{p}_b}{m_a m_b}-\frac{(\bm{n}_{ab}\cdot\bm{p}_a)(\bm{n}_{ab}\cdot\bm{p}_b)}{m_a 
  m_b}\right] \nonumber\\
  &+\frac{1}{2}\sum_a\sum_{b\neq a}\sum_{c\neq a}\frac{G^2 m_a m_b m_c}{r_{ab} r_{ac}}\,, \label{eq:H1PN} \\
  c^2 H_{\rm SO}&=\sum_a\sum_{b\neq a}\frac{Gm_a m_b}{2r_{ab}^2}\bm{n}_{ab}\cdot\left[\frac{\bm{S}_a}{m_a}
  \times\left(\frac{2\bm{p}_b}{m_b}-\frac{3\bm{p}_a}{2m_a}\right)\right.\nonumber\\
  &\left.-\frac{\bm{S}_b}{m_b}\times\left(\frac{2\bm{p}_a}{m_a}-\frac{3\bm{p}_b}{2m_b}\right)\right]\,.
\end{align}
In the above equations, $\bm{p}_a$ is the momentum of body $a$ and $p_a=|\bm{p}_a|$; $\bm{x}_a$ is the position of body $a$ referred to the barycenter; $\bm{r}_{ab}=\bm{x}_a-\bm{x}_b$, and we denote $r_{ab}=|\bm{r}_{ab}|$, $\bm{n}_{ab}=\bm{r}_{ab}/r_{ab}$; $\bm{S}_a$ is the spin of body $a$. For both of the two BHs and the pulsar, we use the dimensionless spin $\chi_a$ defined by $\chi_a=c |\bm{S}_a|/Gm_a^2$ as the parameter. In our timing model, the spins of the three bodies are treated as constants~\cite{Wex:1995pjg}, as the spin precession timescale is much longer than the observation time span in such a system~\cite{Hu:2026zcb}. The summation of $a$, $b$, and $c$ is over the SMBH, IMBH, and pulsar. When we write them out explicitly, we use subscripts or superscripts $S$, $I$, and $P$ respectively. Note that due to the non-linearity of GR, there is a three-body interaction term in the 1\,PN Hamiltonian [the last term in Eq.~(\ref{eq:H1PN})].

We directly use the canonical equations of motion given by 
\begin{equation}
  \dot{\bm{x}}_a=\frac{\partial H}{\partial \bm{p}_a}\,,\quad\quad   \frac{\dot{\bm{p}}_a}{m_a}=-\frac{1}{m_a}\frac{\partial H}
  {\partial \bm{x}_a}\,.
\end{equation}
To be compatible with the cases where the system contains a test-body (for example, we take the pulsar to be massless in later calculations), we in fact solve $\bm{x}_a$ and $\tilde{\bm{p}}_a=\bm{p}_a/m_a$. 

The initial condition of the numerical integration is given at $t=0$, where $t$ is the 
coordinate time. We use a bi-Keplerian model and osculating elements to describe the 
system~\cite{poisson_will_2014}. The bi-Keplerian model was also used to describe the triple-system~\cite{Voisin:2020lqi}. In this model, the pulsar and the SMBH first form an inner binary, and the binary orbit is described by a set of orbital parameters. Then this inner binary and the IMBH form an outer binary, which is described by another set of orbital elements. In the limit that $m_{\rm SMBH}\gg m_{\rm IMBH}$ and $m_{\rm SMBH}\gg m_{\rm PSR}$, these two sets of parameters are the orbital elements of the pulsar and IMBH orbiting around the central SMBH. Therefore in this paper we denote these parameters as
\begin{align}
  \Theta_{P}&= \Big\{P_b^{P},e_{P}, \omega_{P}, \Omega_{P}, i_{P},(f_0)^{P} \Big\}\,,\\
  \Theta_{I}&=\Big\{P_b^{I}\,,e_{I}\,, \omega_{I}\,, \Omega_{I}\,, i_{I}\,,(f_0)^{I} \Big\}\,,
\end{align} 
where the parameters have their usual meanings: $P_b$ is the orbital period; $e$ is the orbital eccentricity; $\omega$ is the longitude of the pericenter; $\Omega$ is the longitude of the ascending node; $i$ is the inclination angle; and $f_0$ is the initial true anomaly.

We shall note that, in pulsar timing, people are used to describe the whole system in the 
harmonic coordinates. The gauge choice affects some formulas, like the R\"{o}mer delay and 
Shapiro delay introduced in the next subsection. The Hamiltonian formulation is written 
in the ADM gauge, and the coordinate transformation to harmonic gauge is given by 
Damour and Sch\"{a}fer~\cite{Damour:1988mr}. Note that this transformation starts at the 2\,PN level, so we ignored it here. In principle, if one wants to build a more realistic timing model, $H_{\rm 2PN}$ and this transformation should be taken into account.

\subsection{Light propagation}\label{subsec:light propagation}

To relate the pulsar's proper rotation to the TOAs of the pulsar's pulse observed at  telescopes on the Earth, one needs to consider the various time delays caused by the light propagation~\cite{Damour:1986}. As mentioned, we only discuss the time delays related to the three-body system, which means that we consider the TOAs at the Solar system barycenter at infinite observing frequency.

The proper rotation of the pulsar, which is related to the pulse number $N$, is described by
\begin{equation}
  N=N_0+\nu T+\frac{1}{2}\dot{\nu}T^2+\cdots\,,
\end{equation}
where $N_0$ is an initial phase; $\nu$ is the spin frequency of the pulsar and $\dot{\nu}$ is the spin-down rate, both of which are measured with respect to $T$, the proper time in the inertial frame of the pulsar. One may regard $T$ as the emission time of the $N$-th pulse.  

Given the pulse emission time $T$, one first needs to translate it to the corresponding coordinate time, which is given by
\begin{equation}
  \frac{{{\rm d}T}}{{\rm d}t}=1-\frac{U}{c^2}-\frac{v^2}{2c^2}+\cdots\,,
\end{equation}
where
\begin{equation}\label{eq:U_E}
  U=\frac{Gm_{S}}{r_{SP}}+\frac{G m_{I}}{r_{IP}}\,,
\end{equation}
is the (negative) Newtonian potential felt by the pulsar, and $v=|{\rm d}\bm{x}_{P}/{\rm d}t|$ is the coordinate velocity of the pulsar. The difference between $T$ and $t$ gives the so-called Einstein delay~\cite{Blandford:1976ApJ} 
\begin{equation}
  \Delta_{\rm E}=t-T\,.
\end{equation}
In a binary system, one can scale the proper time $T$ to fully absorb a linear trend in $\Delta_{\rm E}$. However, for the three-body system we considered here, there is no  analytical expression of the scaling factor. Here we simply use
\begin{equation}\label{eq:Einstein delay}
  \frac{{\rm d}\Delta_{\rm E}}{{\rm d}t}=1-\frac{1-U/c^2-v^2/2c^2}{1-3G(m_{S}+m_{P})/2a_{P}c^2
  -Gm_{I}/a_{I}c^2}\,,
\end{equation}
to calculate the Einstein delay, with $a$ the semi-major axis of the orbit. It shall have captured the leading-order contribution, and higher-order correction can be added when necessary. When we display the Einstein delay, we always further subtract a linear dependence on $t$ from it. We shall note that the scaling is not necessary for the application of the timing model as it is degenerate with the rescaling of $\nu$ and $\dot{\nu}$~\cite{Damour:1986}.

The pulses emitted by the pulsar propagate to the Earth. The contribution from the three-body system is included in the R\"{o}mer delay $\Delta_{\rm R}$ and the Shapiro delay $\Delta_{\rm S}$~\cite{Shapiro:1964uw}. The separation of these two delays is gauge-dependent. In the harmonic gauge, one has
\begin{align}
  \Delta_{\rm R}&=\frac{z_{P}}{c}\,,\\
  \Delta_{\rm S}&=-\frac{2Gm_{S}}{c^3}\ln(r_{PS}-z_{PS})
  -\frac{2Gm_{I}}{c^3}\ln(r_{PI}-z_{PI})\,,
\end{align}
where $z=\bm{x}\cdot\bm{K}_0$ with $\bm{K}_0$ the line of sight direction pointing from the Earth to the GC.

For the purpose of estimating the detectability of the IMBH, in the timing model we only include the leading-order effects introduced above. For a realistic timing model, one should include higher-order effects both in the orbital motion and light propagation until the timing model is compatible with the observation precision (see e.g., Ref.~\cite{Hu:2026zcb}).

Finally, we shall note that our timing model constructed here is very similar in the spirit to the numerical timing model \texttt{NUTIMO}~\cite{Voisin:2020lqi} built for the triple-pulsar system, except that we use the canonical equations of motion and take into account the spin-orbit interaction. In fact, we expect that our timing model can also apply to the triple-pulsar system after some modification. We will explore this aspect in future studies.

\section{Timing residuals}\label{sec:time delays}

In the scenario we consider, unique features in the timing residuals can indicate the existence of an IMBH. Therefore, in this section, we give various examples of the timing residuals caused by the additional IMBH. The largest timing residuals that come from the change in the pulsar's orbital motion caused by the Newtonian gravity of the IMBH are given at the end of this section.

Due to the small amplitudes of the various effects we will discuss compared to the change in the R\"{o}mer delay 
casued by the Netwonian gravity of the IMBH, these effects contribute little to the parameter estimation results shown 
in Sec.~\ref{sec:PE}. However, the effects discussed in this section are still particularly important in the sense that, some of 
them can indicate the existence of the IMBH without finding a fully phase-connected timing solution, and some of them provide 
the possibility to perform unique tests of gravity. 

\subsection{The three-body system}

The configuration of the three-body system can be rather complex. Here we choose $6=2\times 3$ cases as concrete examples. We fix the pulsar's orbit and the parameters of the SMBH while changing the IMBH parameters. The pulsar's orbital parameters are chosen to be
\begin{align}
  P_b^{P}&=0.5\,{\rm yr}\,,\quad e_{P}=0.8\,,\quad \Omega_{P}=0\,,\\
  \omega_{P}&=\frac{5\pi}{7}\,,\quad i_{P}=\frac{\pi}{5}\,,\quad (f_0)^{P}=-\frac{3\pi}{4}\,.
\end{align}
Due to a rotation symmetry around the line of sight direction in the timing observation, we set $\Omega_{P}=0$ as a reference direction. While treating the pulsar as a test particle, for the SMBH we use the following values for its mass and spin,
\begin{align}
  m_{S}&=4.3\times 10^6\,M_\odot\,,\\
  \chi_{S}&=0.6\,,\quad \lambda_{S}=\pi/6\,,\quad \eta_{S}=5\pi/9\,,
\end{align}
where $\lambda_{S}$ and $\eta_{S}$ give the direction of the spin.

For the IMBH, we first consider that it can have different masses and semi-major axes. Constrained by the S0-2 observation~\cite{Will:2023nlt, GRAVITY:2023met}, for an IMBH with $m_{\rm I}\gtrsim 10^3 \,M_\odot$, its semi-major axis is likely larger than $10^3\,{\rm AU}$. For an IMBH with $m_{\rm I}\sim 10^2\,M_\odot$, the current constraint on its semi-major axis is rather weak. Therefore, we consider two combinations of the IMBH's mass and its orbital period
\begin{equation}
  m_{I}=10^3\,M_\odot\,,\ P_b^I=20\,{\rm yr}\,, \label{eq:IMBH:mass:Pb:1}
\end{equation}
or 
\begin{equation}
  m_{I}=10^2\,M_\odot\,,\ P_b^I=3\,{\rm yr}\,. \label{eq:IMBH:mass:Pb:2}
\end{equation}

We further consider three different inclinations of the IMBH orbit, and denote them by: Case (I) $i_{I}=\pi/2$; Case~(II) $i_{I}=0$; and Case (III) $i_{I}=i_{P}$. Other orbital parameters of the IMBH are fixed as
\begin{equation}
  e_{I}=0.5\,,\quad \Omega_{I}=0\,,\quad \omega_{I}=\frac{\pi}{3}\,,\quad (f_0)^{I}=-\frac{17\pi}{18}\,. \label{eq:IMBH:orbit}
\end{equation}
We choose $\Omega_{I}=0$ so that in Case (III) the orbits of the pulsar and the IMBH are in the same plane. We choose $\omega_{I}+(f_0)^{I}=-17\pi/18$ so that in Case (I) the IMBH will move across right in front of the SMBH during the observational time span. In Eq.~(\ref{eq:IMBH:orbit}), $\omega_{I}=\pi/3$ was rather arbitrarily chosen.

Finally, though we introduce the spin-orbit coupling between each pair of two bodies, only the spin of the SMBH is relevant here~\cite{Liu:2011ae}. Ignoring the spin of the pulsar, the spin-orbit coupling force caused by the IMBH on the pulsar is at the order of $\sim  \chi_I \big|4\bm{v}_P-3\bm{v}_I \big|  \cdot G^2  m_I^2 /c^3r_{PI}^3$. Even considering an IMBH with a relatively tight orbit so that $r_{\rm PI}\sim r_P$, this force is about a factor of $(m_I/m_S)^2\lesssim 10^{-6}$ smaller than the spin-orbit coupling force caused by the SMBH and can be safely ignored considering the studies of the spin effect caused by the SMBH~\cite{Liu:2011ae, Hu:2026zcb}. Therefore, we ignore the spin of the IMBH in this work. In numerical calculation, we set $\chi_{I}=0$.

\subsection{Shapiro delay}

Shapiro delay in pulsar timing provides a unique way of measuring the companion mass, as its amplitude is proportional to the mass of the pulsar's companion, while in the case of a nearly edge-on configuration it has a sharp shape that is clearly distinguishable from other effects~\cite{Liu:2011ae}.  In normal binary pulsar systems, the Shapiro delay is mainly detectable when the system has an inclination angle close to $90^\circ$ due to the small mass of the companion. In contrast, the IMBH might cause a relatively large Shapiro delay even for systems with general inclination angles, considering its large mass. Further, spike structures in timing residuals can be noticeable evidence for the presence of an IMBH passing in front of the pulsar.

\begin{figure}[H]
	\begin{center}
		\includegraphics[width=8.5cm]{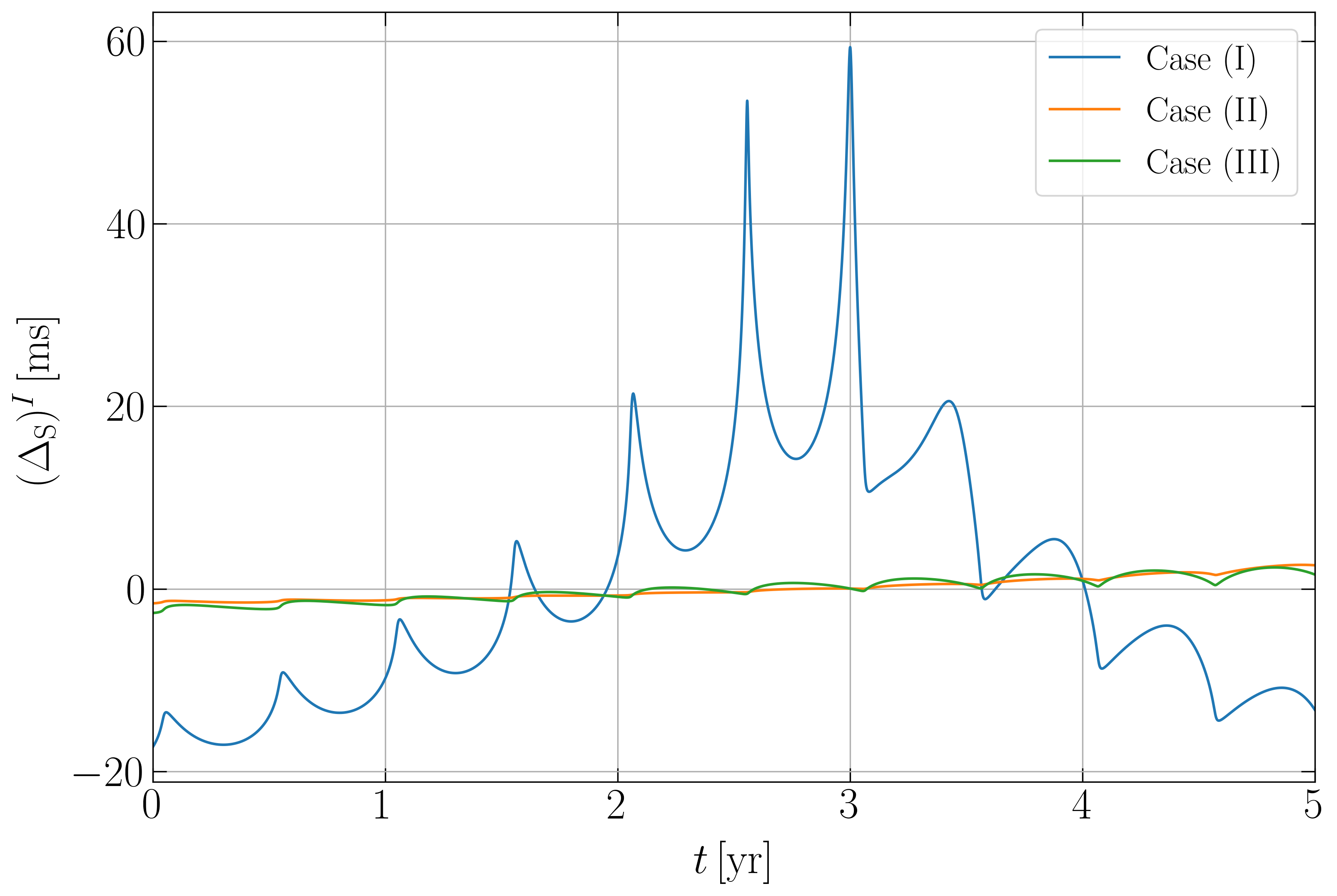}
	\caption{Examples for the Shapiro delay for the cases with $m_{I}=10^3\,M_\odot$ and $P_b^{I}=20\,{\rm yr}$ in Eq.~(\ref{eq:IMBH:mass:Pb:1}). A constant term is removed from the time delays.\label{fig:Shapiro_Pb20}}
	\end{center}
\end{figure}

\begin{figure}[H]
	\begin{center}
		\includegraphics[width=8.5cm]{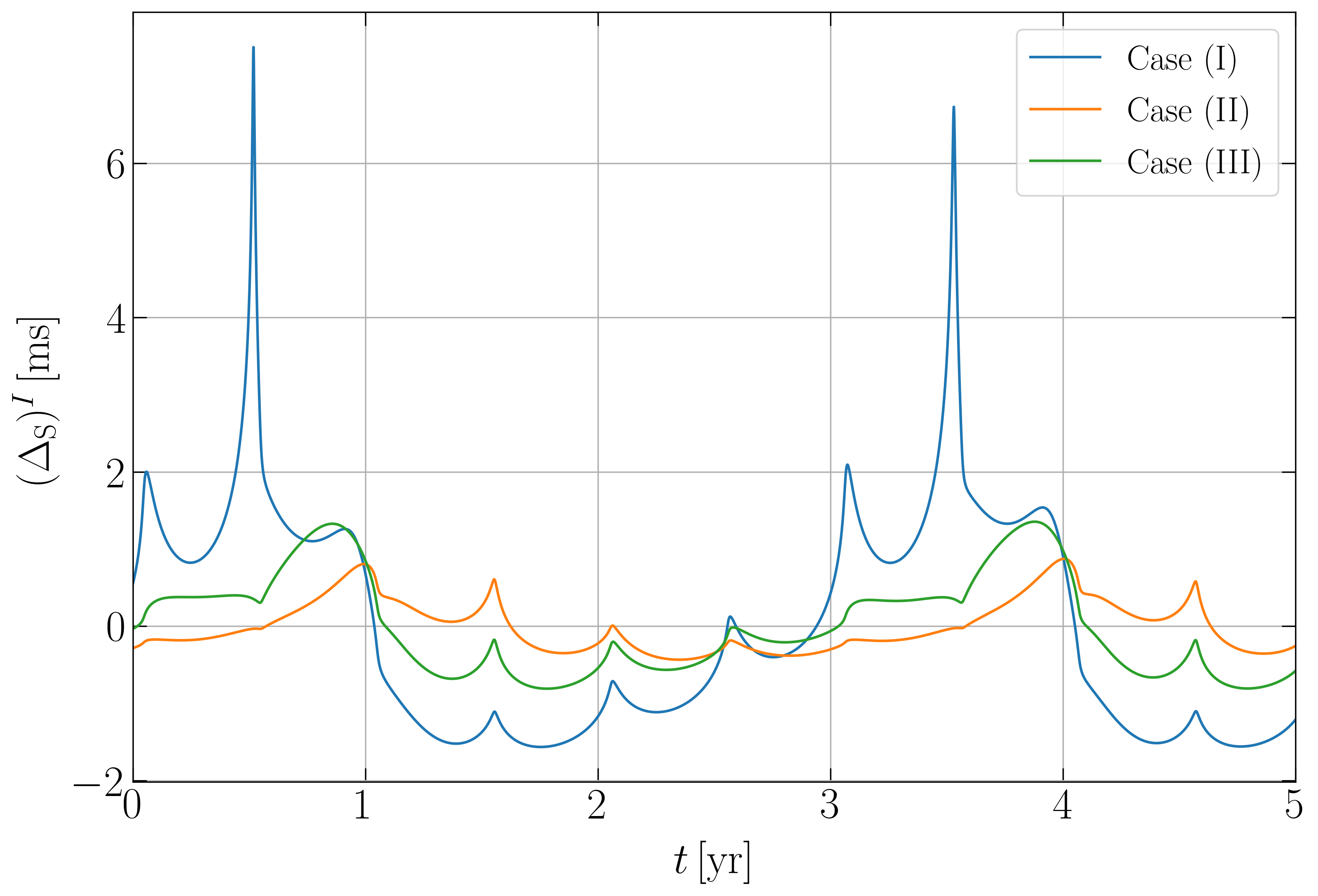}
	\caption{ Examples for the Shapiro delay for the cases with $m_{I}=10^2\,M_\odot$ and $P_b^{I}=3\,{\rm yr}$ in Eq.~(\ref{eq:IMBH:mass:Pb:2}). \label{fig:Shapiro_Pb3}}
	\end{center}
\end{figure}

In Fig.~\ref{fig:Shapiro_Pb20} and Fig.~\ref{fig:Shapiro_Pb3} we show the Shapiro delay contribution caused by the IMBH, 
\begin{equation}
	(\Delta_{\rm S})^{I}=-\frac{2Gm_{I}}{c^3}\ln(r_{PI}-z_{PI})\,,
\end{equation}
for all six examples. Except for the very special Case (I) in Fig.~\ref{fig:Shapiro_Pb20}, for most cases the Shapiro delay caused by the IMBH has an amplitude of about 1--2 milliseconds. Different from the Shapiro delay in a normal binary system, which has a periodic spike structure, the Shapiro delay in the pulsar-SMBH-IMBH system shows a more complex time dependence. The orbital motion of the pulsar and the IMBH both leave imprints, so the time delay is modulated by two periods.

In general, for an IMBH in orbit with an orbital period significantly larger than the pulsar orbit, one can expand the Shapiro delay caused by the IMBH as
\begin{equation}
  \left(\Delta_{\rm S}\right)^{I} \approx -\frac{2Gm_{I}}{c^3}\left[\ln(r_I+z_I)-\frac{\bm{n}_{I}\cdot \bm{x}_{P}+z_{P}}{r_{I}+z_{I}}\right]\,,
\end{equation}
where the factor $r_I+z_I$ provides a long-time modulation related to the orbital motion of the IMBH, and the factor $\bm{n}_{I}\cdot \bm{x}_{P}+z_{P}$ mainly changes at the timescale of the pulsar's orbital motion. For general cases with $r_I+z_I\sim a_I$, the amplitude of the Shapiro delay is about $2Gm_Ia_P/a_Ic^3$. For the system considered in Fig.~\ref{fig:Shapiro_Pb20}, this gives  $\sim 1\,{\rm ms}$ as expected. The very special Case (I) in Fig.~\ref{fig:Shapiro_Pb20} is designed to have $(r_I+z_I)/a_I\sim 0$ during the observation time span, so that it largely amplifies the Shapiro delay. 

For an IMBH with a smaller orbital period, the amplitude and shape of the Shapiro delay caused by the IMBH are more complex. For an observation time span that is long enough, one can see the modulation caused by both the IMBH's and the pulsar's orbital motions, as shown in Fig.~\ref{fig:Shapiro_Pb3}. For a general orbital configuration and an IMBH mass allowed by current constraints, the amplitude of the Shapiro delay is at the order of 1--10$\,{\rm ms}$.

\subsection{Einstein delay}

The Einstein delay is a combination of the gravitational redshift and special-relativistic time-dilation effects. Therefore, as the IMBH also affects the pulsar's orbital motion, there is no clear separation of the Einstein delay caused by the IMBH and the SMBH, unlike in the Shapiro delay. Nevertheless, for the Einstein delay, one can illustrate the IMBH effect by considering the additional gravitational redshift in the original pulsar orbit. Numerically, we can integrate the equations of motion with $m_I=0$ but set $m_I$ to be the desired mass in Eq.~(\ref{eq:U_E}). 

\begin{figure}[H]
	\begin{center}
		\includegraphics[width=8.5cm]{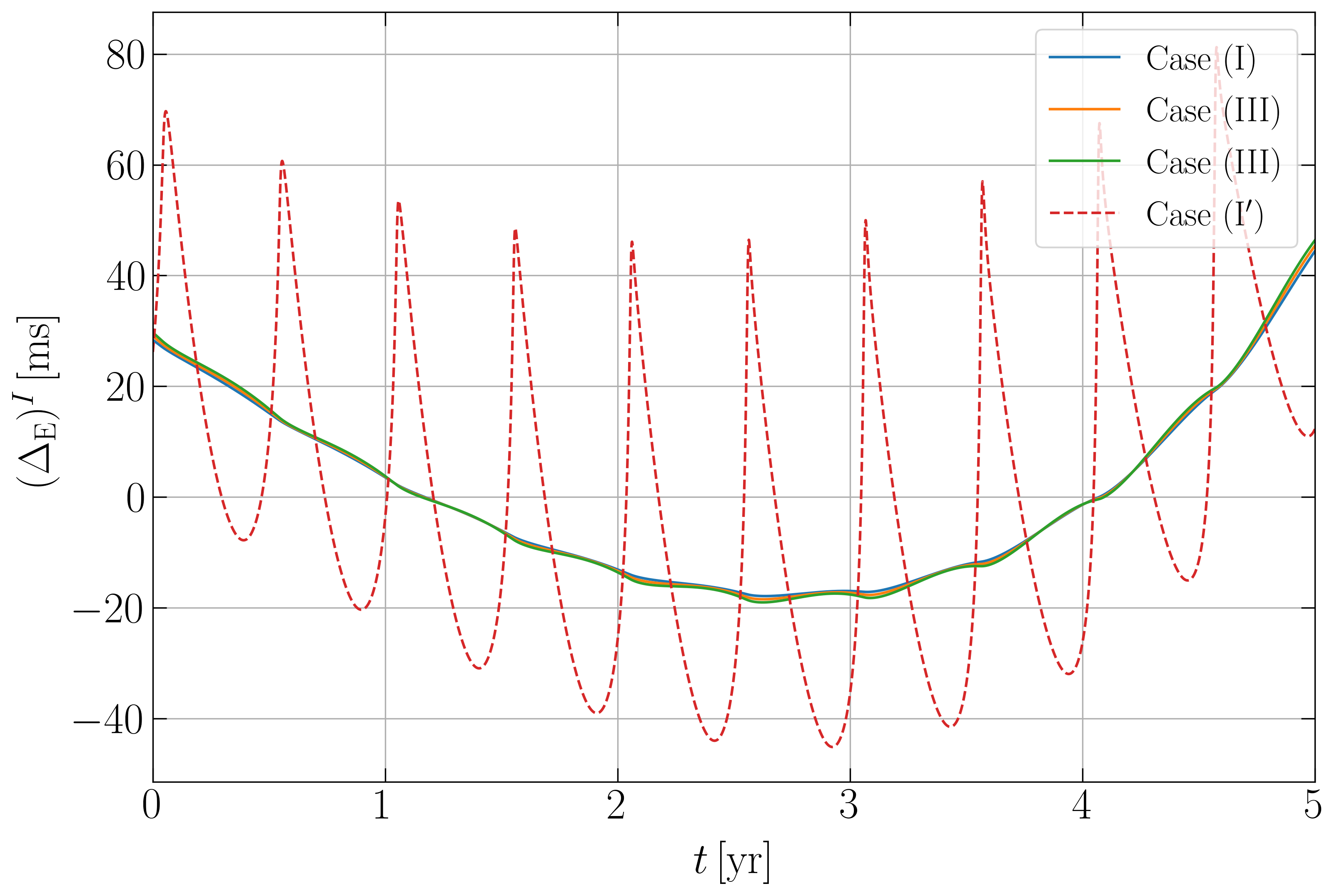}
	\caption{Examples for the Einstein delays related to the additional gravitational redshift caused by the IMBH, with $m_{I}=10^3\,M_\odot$ and $P_b^{I}=20\,{\rm yr}$ in Eq.~(\ref{eq:IMBH:mass:Pb:1}). We remove the linear dependence on $t$ in the plot. Case (I$'$) shows the result of directly comparing the Einstein delays from calculations with and without the IMBH. \label{fig:Einstein_Pb20}}
	\end{center}
\end{figure}

In Fig.~\ref{fig:Einstein_Pb20}, we show the Einstein delay related to the gravitational redshift caused by the IMBH for cases with $m_{I}=10^3\,M_\odot$ and $P_b^{I}=20\,{\rm yr}$. The calculations for the solid lines only take into account the additional gravitational redshift as discussed above. In addition, for Case (I), we show the direct difference between the Einstein delays from calculations with and without the IMBH, including the difference in the pulsar's orbital motion [labeled with Case (I$'$)]. Note that, strictly speaking, such a calculation is improper if the IMBH changes the pulsar's orbital motion significantly. In the figure, the linear trends of $t$ are subtracted from the Einstein delay, as it is equivalent to a rescaling of the proper time $T$ of the pulsar. 

In these three cases, the change in the pulsar orbit caused by the IMBH is not significant, so the direct subtraction results shown by Case (I$'$) do not give increasing residuals due to the cumulating orbital phase shift. However, the oscillation pattern in Case (I$'$)  may still come from a small difference in the pulsar orbit from the calculations with and without the IMBH. Nevertheless, for these cases, the Einstein delays related to the gravitational redshift show a consistent trend and amplitude for the non-oscillatory signal with the direct subtraction result, which suggests that they have captured the main feature of the Einstein delay caused by the IMBH.

\begin{figure}[H]
	\begin{center}
		\includegraphics[width=8.5cm]{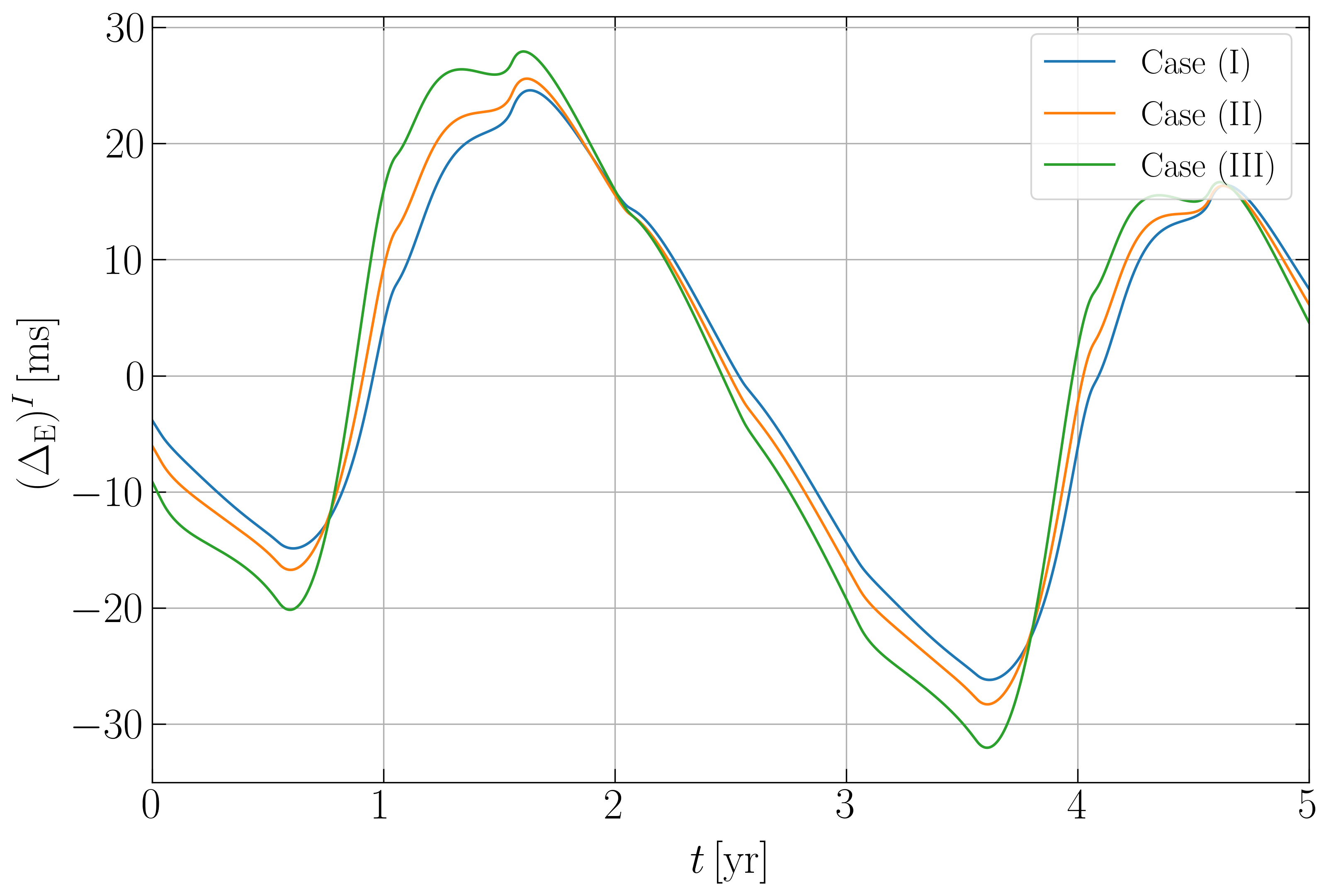}
	\caption{Examples for Einstein delay for cases with $m_{I}=10^2\,M_\odot$ and $P_b^{I}=3\,{\rm yr}$ in Eq.~(\ref{eq:IMBH:mass:Pb:2}). We have removed the linear trend as in Fig.~\ref{fig:Einstein_Pb20}. \label{fig:Einstein_Pb3}}
	\end{center}
\end{figure}

Similarly, in Fig.~\ref{fig:Einstein_Pb3}, we show the Einstein delays related to the gravitational redshift caused by the IMBH for cases with $m_{I}=10^2\,M_\odot$ and $P_b^{I}=3\,{\rm yr}$. In these cases, the results show a period of $3\,{\rm yr}$ as expected. 

\subsection{Three-body interaction}

Due to the non-linear nature of GR, in the 1\,PN Hamiltonian, there is a three-body interacting term 
\begin{equation}
  c^2H_{\rm 1PN}^{\rm 3\mbox{-}body}=\frac{G^2 m_{P} m_{I} m_{S}}{r_{PI}r_{IS}r_{SP}}
  (r_{SP}+r_{PI}+r_{IS})\,.
\end{equation}
In the limit $m_{P}\rightarrow 0$, this term still affects the pulsar's motion. Solar system experiment is expected to detect the additional periastron advance of the Mercury caused by the three-body interaction between the Sun, the Mercury, and other planets in the near future~\cite{Will:2018mcj}. It is interesting whether in the pulsar-SMBH-IMBH system one can also measure this term as a new test of GR.

\begin{figure}[H]
	\begin{center}
		\includegraphics[width=8.5cm]{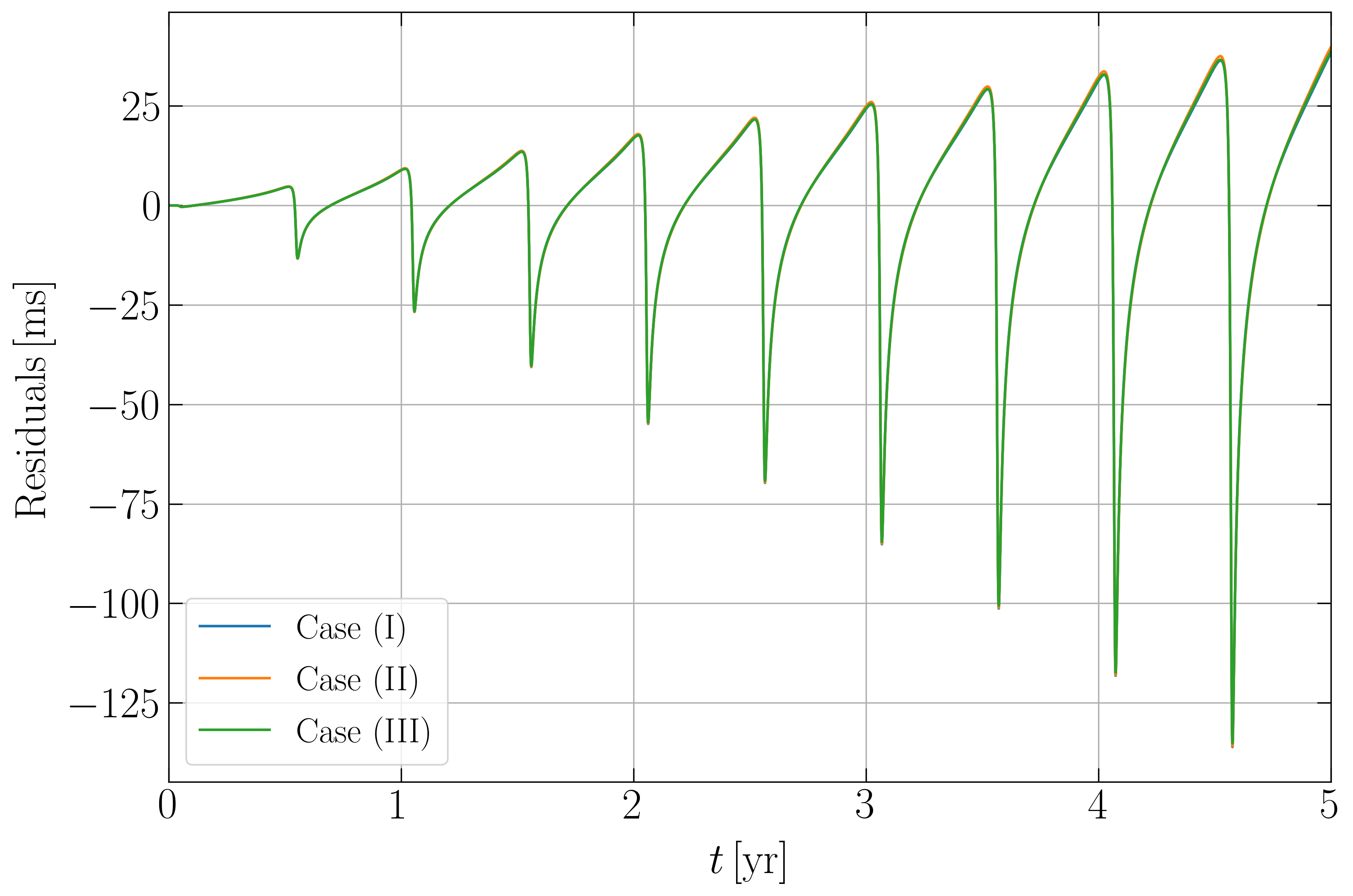}
	\caption{Examples of the timing residuals caused by the three-body interaction term with $m_{I}=10^3\,M_\odot$ and $P_b^{I}=20\,{\rm yr}$ in Eq.~(\ref{eq:IMBH:mass:Pb:1}). The figure shows the pre-fit residuals, which are a direct comparison of the TOAs from systems with $f_3=1$ and $f_3=0$. Results from three cases are largely overlapping. \label{fig:f3_Pb20} }
	\end{center}
\end{figure}

\begin{figure}[H]
	\begin{center}
		\includegraphics[width=8.5cm]{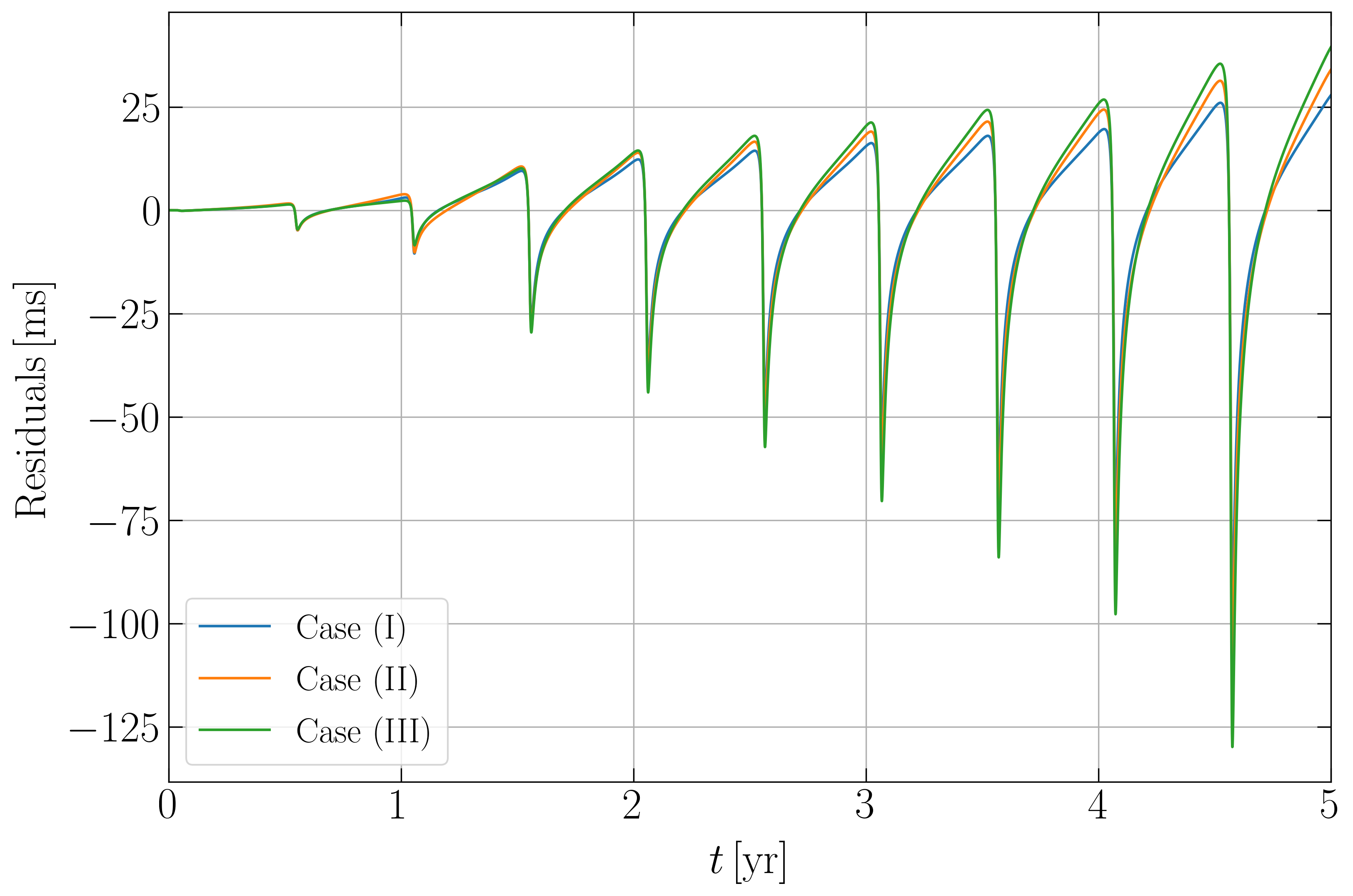}
	\caption{Similar to Fig.~\ref{fig:f3_Pb20} but for an IMBH with $m_{I}=10^2\,M_\odot$ and $P_b^{I}=3\,{\rm yr}$ in Eq.~(\ref{eq:IMBH:mass:Pb:2}). \label{fig:f3_Pb3}}
	\end{center}
\end{figure}

In Fig.~\ref{fig:f3_Pb20} and Fig.~\ref{fig:f3_Pb3}, we present the pre-fit timing residual caused by the three-body interacting term. To achieve this, in the timing model, we introduce an additional factor $f_3$ that is multiplied to this term controlling its amplitude. The pre-fit timing residual in the figures is a direct comparison of the TOAs of the systems with $f_3=1$ (with full three-body interaction) and $f_3=0$ (without three-body interaction).

For all six cases, the pre-fit timing residuals caused by the three-body interaction term show a similar amplitude. For the 5-year time span, the timing residual can cumulate to around $150\,{\rm ms}$ as the three-body interaction introduces secular effects~\cite{Will:2018mcj}. These pre-fit residuals are significantly larger than the assumed timing precision, which is about $1\,{\rm ms}$ or smaller for future observations with the SKA~\cite{Liu:2011ae}. Therefore it is necessary to take the three-body interaction into account in a realistic timing model for pulsar-SMBH-IMBH systems.

\subsection{Post-fit residual}\label{subsec:post-fit}

In the pulsar-SMBH-IMBH systems, the main contribution of timing residuals caused by the IMBH  comes from the change of the pulsar's orbital motion due to the Newtonian gravity of the IMBH. This effect is mixed with all the other effects shown before. Here, we compare the full timing models with and without the IMBH through the post-fit timing residuals. We perform a least-square fitting to the TOAs generated by the timing model that includes the IMBH effects with a simpler timing model that only considers the pulsar and the SMBH. The remaining difference in the TOAs then gives the post-fit timing residual. In timing observations, large post-fit timing residuals with clear structures often suggest that important physical effects are missed in the timing model.

\begin{figure}[H]
	\begin{center}
		\includegraphics[width=8.5cm]{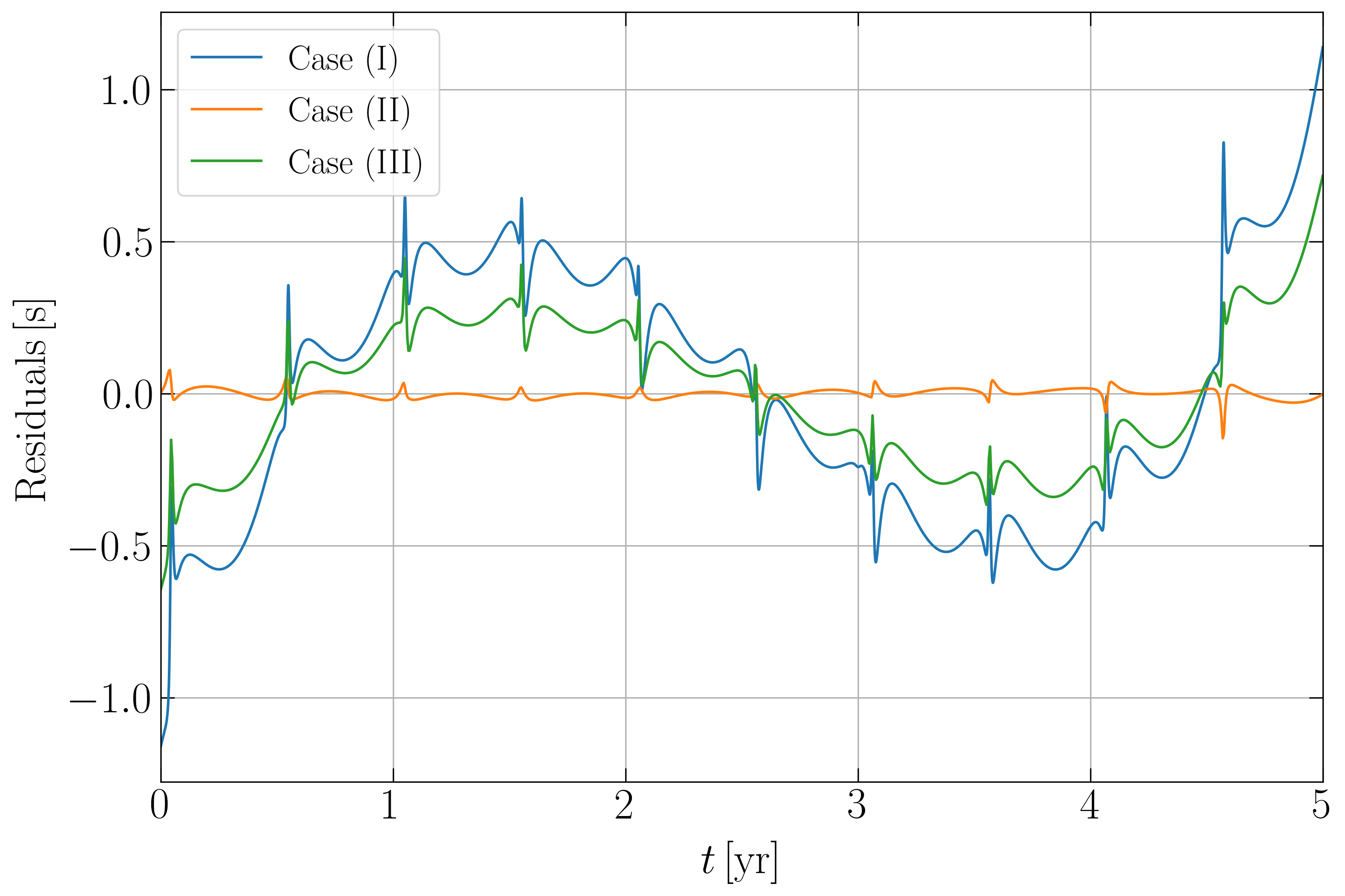}
	\caption{Examples of post-fit timing residuals for systems with $m_{I}=10^3\,M_\odot$ and $P_b^{I}=20\,{\rm yr}$ in Eq.~(\ref{eq:IMBH:mass:Pb:1}).  \label{fig:postfit_Pb20}}
	\end{center}
\end{figure}

\begin{figure}[H]
	\begin{center}
		\includegraphics[width=8.5cm]{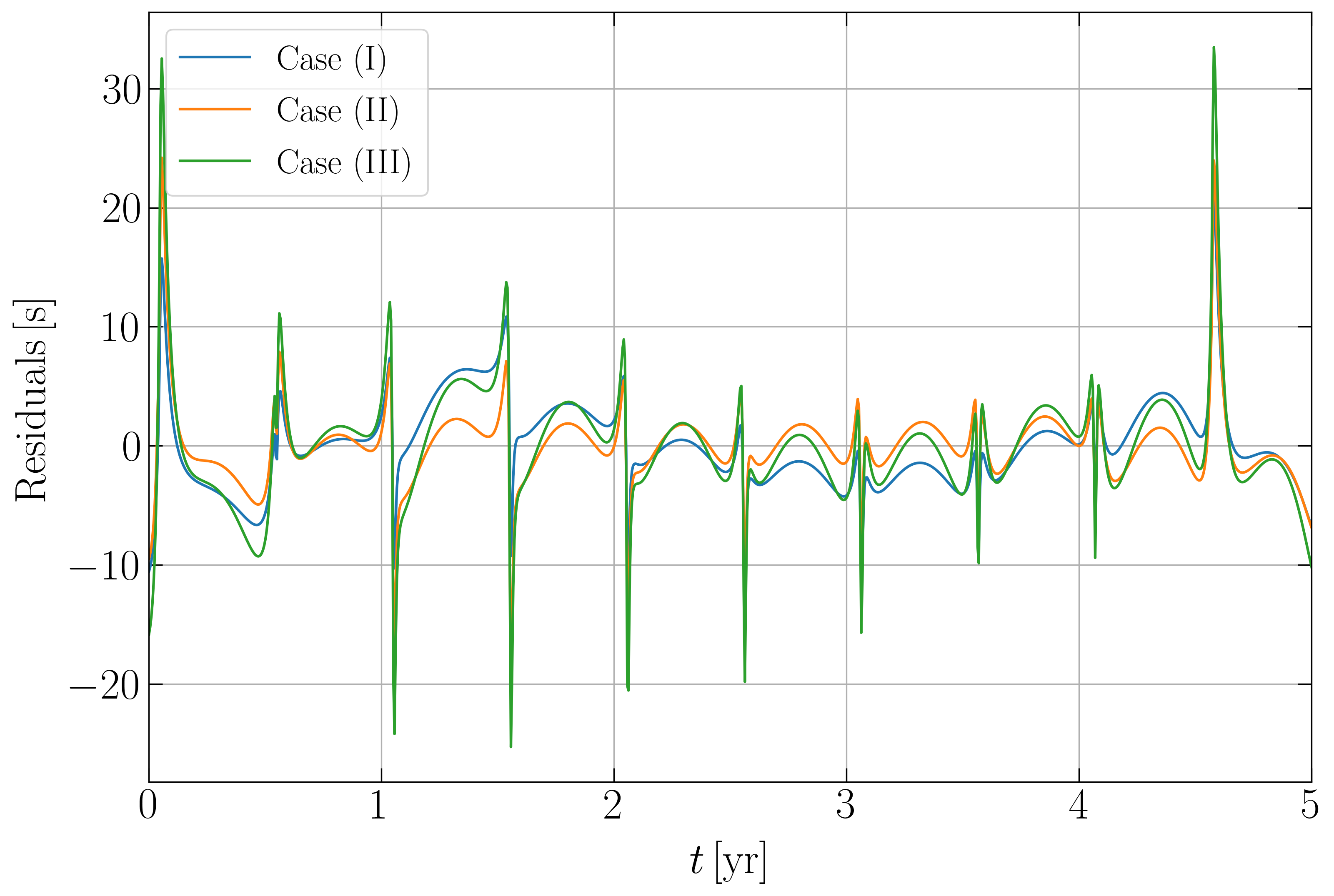}
	\caption{Examples of post-fit timing residuals for systems with  $m_{I}=10^2\,M_\odot$ and $P_b^{I}=3\,{\rm yr}$ in Eq.~(\ref{eq:IMBH:mass:Pb:2}). \label{fig:postfit_Pb3}}
	\end{center}
\end{figure}

In Fig.~\ref{fig:postfit_Pb20} and Fig.~\ref{fig:postfit_Pb3}, we show the post-fit 
timing residuals of all six examples we considered. We can see that, though the mass 
of the IMBH is smaller for cases with $m_{I}=10^2\,M_\odot$ and $P_b^{I}=3\,{\rm yr}$ 
in Eq.~(\ref{eq:IMBH:mass:Pb:2}), the post-fit timing residual is significantly larger, 
which is expected as the IMBH affects the pulsar's orbital motion in a more complex 
manner. The IMBH introduces effects with frequencies different from the pulsar's 
orbital motion, and are hard to be absorbed by the simpler pulsar-SMBH timing model. 
Such a large timing residual that is much higher than the timing precision will 
indicate super-high measurement precision of the 
IMBH parameters, as we will discuss in later sections. This large post-fit residual 
also suggests that the pulsar-SMBH system is sensitive to the environment perturbations, 
which might be strong in the GC, and relevant effects must be taken into account in 
real observations~\cite{Hu:2026aez}.

\section{Parameter estimation}\label{sec:PE}

Though we know that directly performing parameter estimation would provide unrealistic results as hinted by the large post-fit timing residuals, we still ignore all other stellar perturbations in this section and study the idealized measurement uncertainties for illustration. In the next section, we will discuss the detectability of an IMBH with more realistic assumptions.

We apply the Fisher matrix analysis to estimate the parameter measurement precision~\cite{Edwards:2006zg}. Assuming a Gaussian timing noise realization in observation, the likelihood function reads
\begin{equation}
	P(\Theta|t^{\rm TOA})\propto \exp\left(-\frac{1}{2\nu^2}\sum_{i=1}^{N_{\rm TOA}}\frac{\big[N_i(\Theta)-N_i(\tilde{\Theta})\big]^2}{\sigma^2_{\rm TOA}}\right)\,,
\end{equation}
where $N_i(\Theta)=N(t^{\rm TOA}_i;\Theta)$ is the predicted rotation number of the $i$-th TOA received at time $t^{\rm TOA}_i$; $\Theta$ denotes the system parameters, and we use $\tilde{\Theta}$ to denote their true values. Here we list the parameters of the pulsar-SMBH-IMBH system, 
\begin{align}
  \Theta&=\Theta_S\cup\Theta_I\cup\Theta_P\,,\\
  \Theta_S&= \big\{m_S,\chi,\lambda,\eta \big\}\,,\\
  \Theta_I&= \Big\{m_I,P_b^I,e^I,\Omega^I,\omega^I,i^I,f_0^I \Big\}\,,\\
  \Theta_P&= \Big\{P_b^P,e^P,\omega^P,i^P,f_0^P,N_0,\nu,\dot{\nu} \Big\}\,.
\end{align}
As discussed before, we treat the pulsar as a test particle and ignore the spin of the IMBH. In our simplified timing model, $\Omega_P$ is not measurable and is taken to be zero as a reference direction. When studying the three-body interaction term, we will further consider an additional parameter $f_3$ which controls the amplitude of the three-body interaction.  

The measurement precisions of the SMBH parameters via pulsar timing were studied in detail in many previous studies~\cite{Liu:2011ae, Psaltis:2015uza, Zhang:2017qbb, Hu:2023ubk, Hu:2026zcb}, and we are not presenting them here as they are not affected too much by the presence of the IMBH. However, we shall mention that the existence of an IMBH can still bias the best-fit value of the SMBH parameters if the timing model treats the IMBH perturbation improperly.

\begin{figure}[H]
	\begin{center}
		\includegraphics[width=8.5cm]{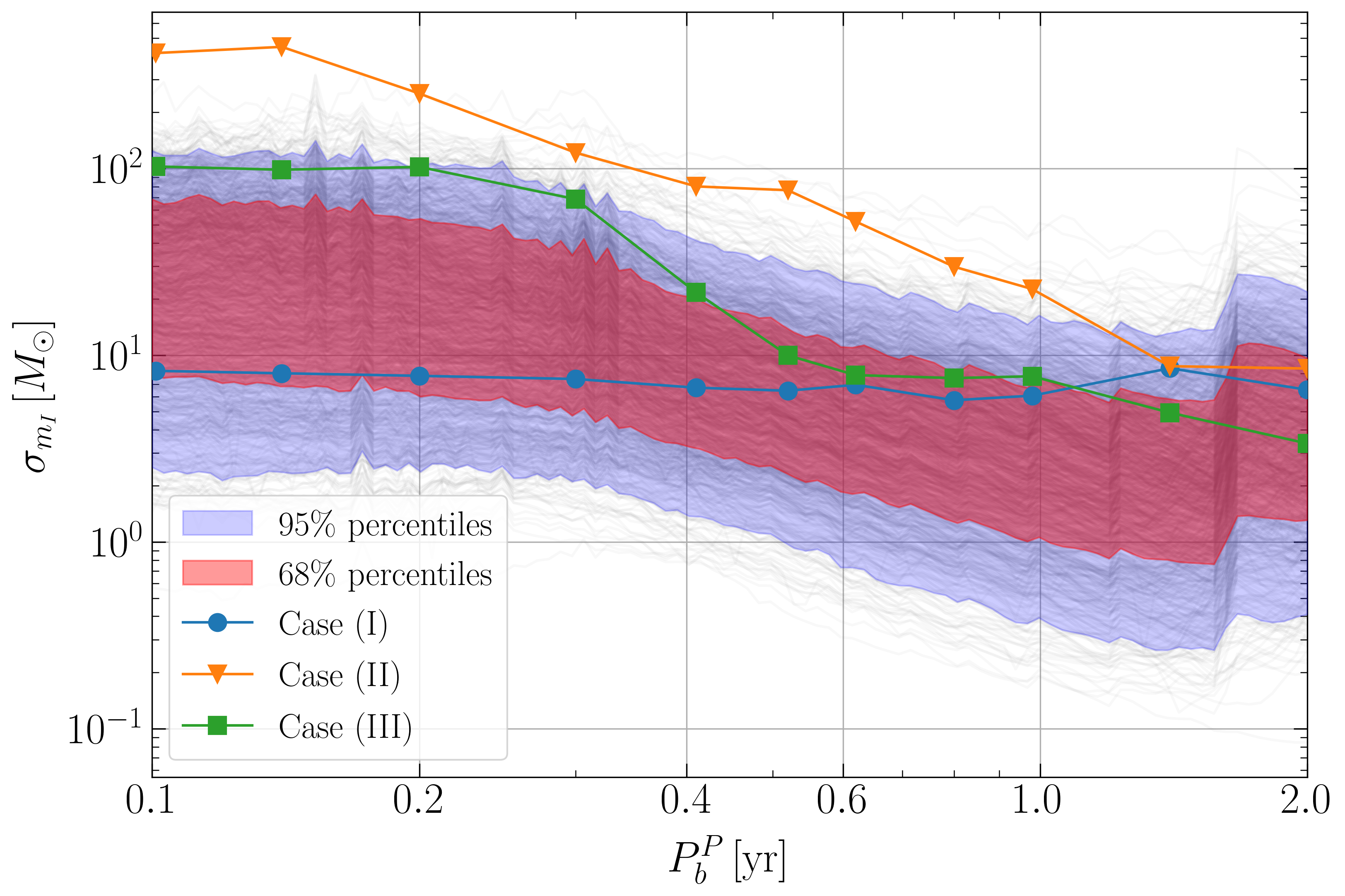}
	\caption{Measurement precision of the IMBH mass as a function of the pulsar's orbital period. Cases shown in this figure have an IMBH with $m_{I}=10^3\,M_\odot$ and $P_b^{I}=20\,{\rm yr}$. The faint gray lines represent the simulations with random IMBH orbital orientations, with the red and blue colors marking the central $68\%$ and $95\%$ percentile regions.} \label{fig:1e3_Pb}
	\end{center}
\end{figure}

\begin{figure}[H]
	\begin{center}
		\includegraphics[width=8.5cm]{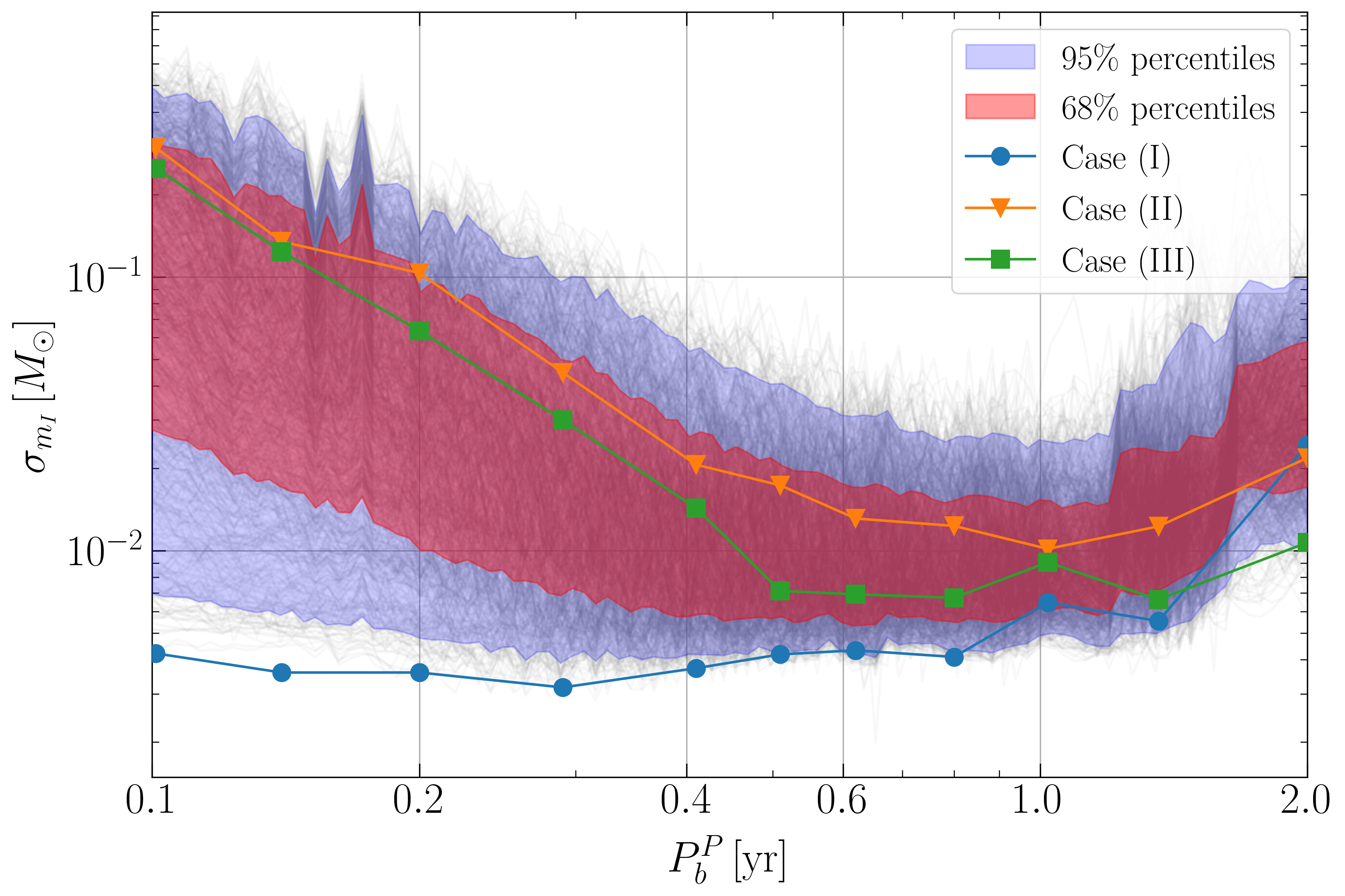}
	\caption{Similar to Fig.~\ref{fig:1e3_Pb} but for cases for an IMBH with $m_{I}=10^2\,M_\odot$ and $P_b^{I}=3\,{\rm yr}$.} \label{fig:1e2_Pb}
	\end{center}
\end{figure}

Here we focus on the estimation of the detectability of the IMBH. We present the measurement precision of the mass of the IMBH for our illustrative cases. Further, to ensure that our results are not sensitive to a particular choice of orbital configuration, we perform $1000$ simulations with random orientation of the IMBH orbits for each choice of the IMBH mass, orbital period, and orbital eccentricity. For the cases with an IMBH in a 20-${\rm yr}$ orbit, the results are shown in Fig.~\ref{fig:1e3_Pb}.  The measurement precision of the IMBH mass becomes better when the pulsar has a larger orbital period, as in general such measurement is more sensitive when the pulsar orbit and the IMBH orbit have comparable sizes. Though Case (I) is relatively special, a measurement precision of about $10\, M_\odot$ suggests that even the perturbations caused by the S-stars might be measurable via pulsar timing~\cite{Gillessen:2017jxc}. In other words, for pulsars with orbital periods in the order of years, perturbations from the known S-stars are already non-negligible.

The estimated precisions shown in Fig.~\ref{fig:1e2_Pb} for an IMBH with $m_{I}=10^2\,M_\odot$ and $P_b^{I}=3\,{\rm yr}$ are clearly unrealistic. The results suggest that any object with a mass larger than 0.01--0.1$\,M_\odot$ would cause observable effects on the timing observation. Considering the stellar cluster around the GC SMBH~\cite{Peebles:1972}, there could be a large population of such objects, consisting of low-mass main-sequence stars, BHs, neutron stars, and white dwarfs, and they can spoil the measurement. In the sense that one might still be able to distinguish the IMBH signal,  it is unrealistic to have such a high measurement precision. In another study, we have discussed the effect of the granular mass perturbation from stellar objects on the timing observation of the pulsar-SMBH system~\cite{Hu:2026aez}. These results can be compared to get a qualitative picture concerning different types of perturbations.

\begin{figure}[H]
	\begin{center}
		\includegraphics[width=8.5cm]{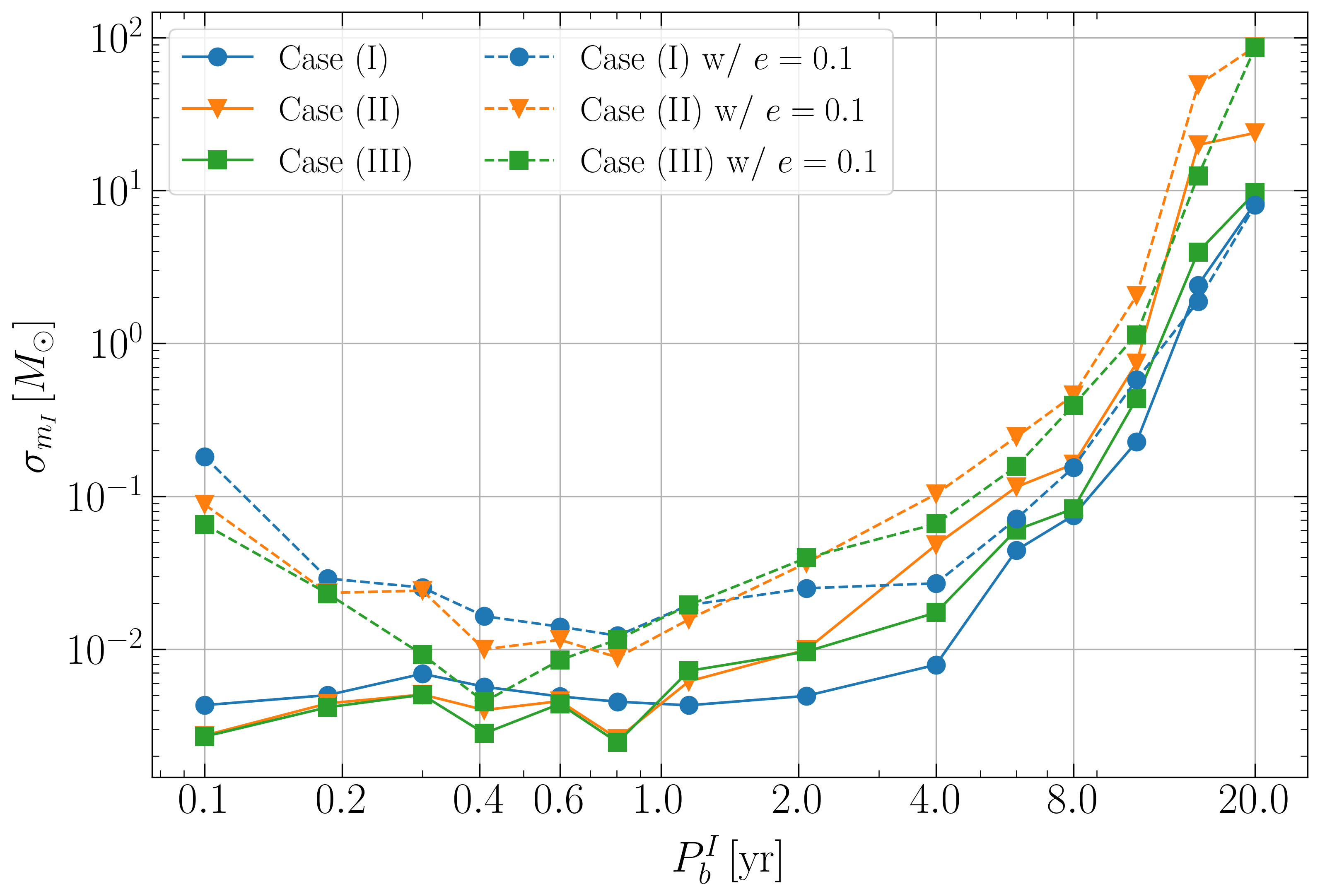}
	\caption{ Measurement precision of the IMBH mass as a function of the IMBH orbital period. The pulsar is fixed to a 0.5-yr orbit. The solid lines show the three cases defined before, except that the IMBH mass is now fixed to be $10^2\,M_\odot$ and its orbital period is not fixed. The dashed lines show cases for the pulsar with an orbital eccentricity $e^P=0.1$.  \label{fig:MI_PBI}  }
	\end{center}
\end{figure}

In Fig.~\ref{fig:MI_PBI}, we show the measurement precision of the IMBH mass as a function of the orbital period of the IMBH while keeping the pulsar's orbital period to $0.5\,{\rm yr}$. Here we keep the mass of the IMBH to $10^{2}\,M_\odot$.
The mass of the IMBH could affect the parameter estimation result of $\sigma_{m_I}$. 
Nevertheless, as long as the mass of the IMBH is small enough so that it does not change 
the pulsar's orbit significantly, one would expect that $\sigma_{m_I}$ only depends on 
$m_I$ mildly. In Fig.~\ref{fig:MI_PBI}, dashed lines represent cases where the pulsar has a more circular orbit, $e^P=0.1$. As expected, compared to a more eccentric orbit shown by the solid lines, a circular orbit has a worse measurement precision of the IMBH with a small orbital period. These examples clearly illustrate that such a pulsar-SMBH-IMBH system is more sensitive to an IMBH with a comparable orbit size as the pulsar.

We should note again that the very high measurement precision of the mass of the IMBH shown in this figure is unrealistic. Complex environmental perturbations will spoil such measurements. However, treating the environmental effects as an effective background noise, one would expect that the shape of the curve will roughly be kept even in more realistic cases, while the whole curve may shift upwards to give a worse sensitivity. This argument is consistent with the results shown in Sec.~\ref{sec:detect}.

\begin{figure}[H]
	\begin{center}
		\includegraphics[width=8.5cm]{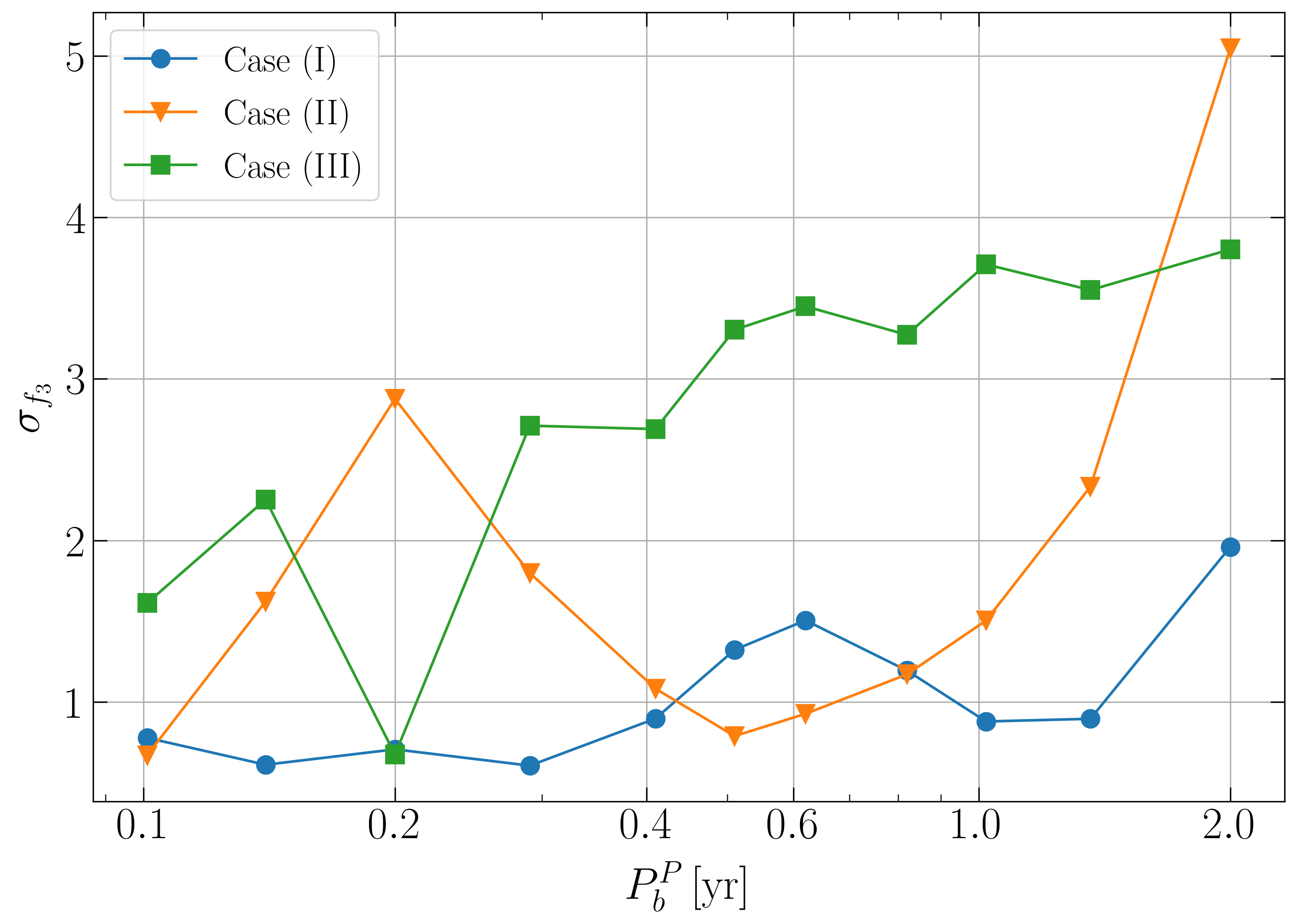}
	\caption{ Measurement precision of the three-body interaction term parametrized by $f_3$. Results in this figure have assumed an IMBH with $m_I=10^3\, M_\odot$ and $P_b^I=20\,{\rm yr}$. \label{fig:f3_1e3} }
	\end{center}
\end{figure}

\begin{figure}[H]
	\begin{center}
		\includegraphics[width=8.5cm]{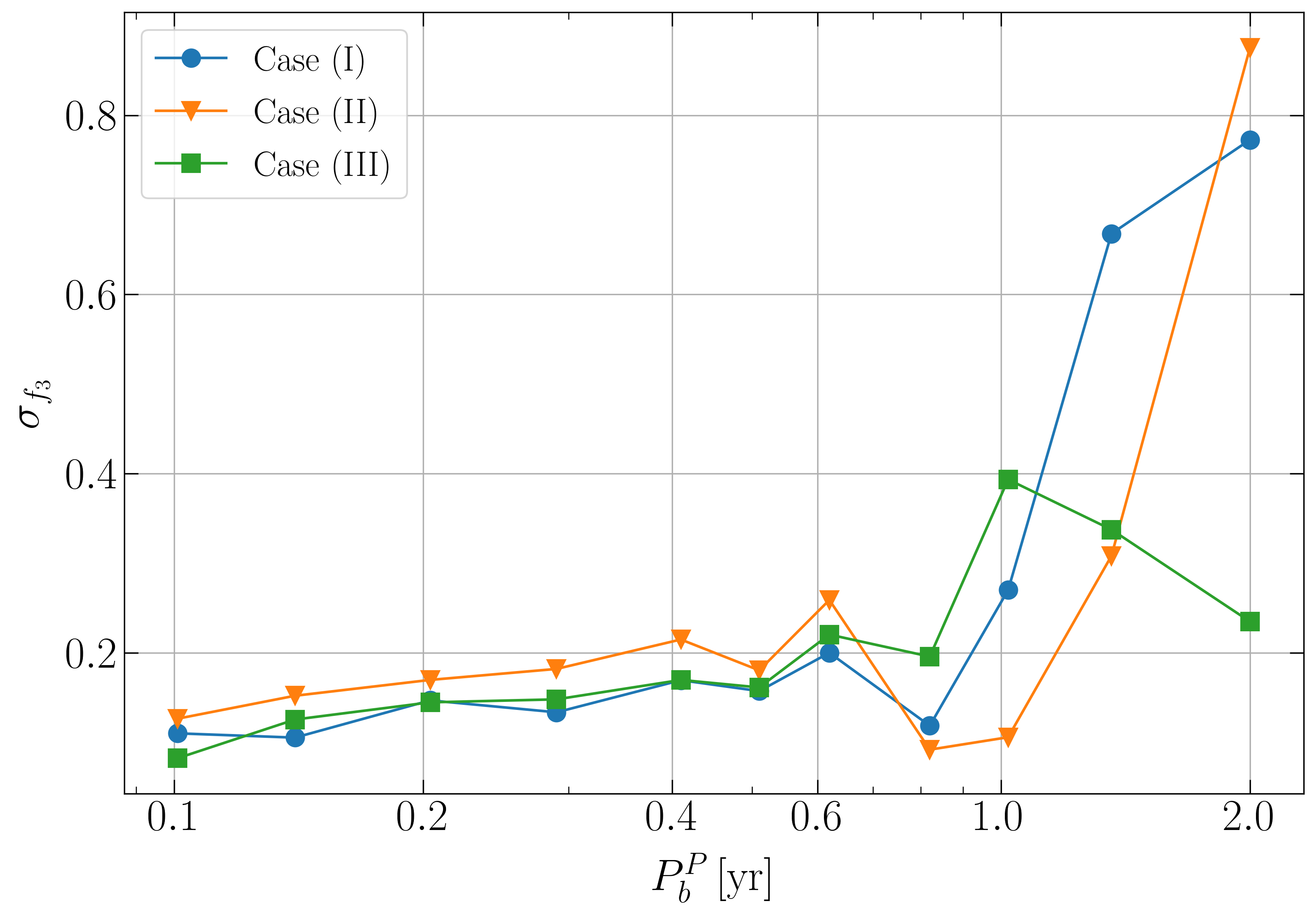}
	\caption{ Similar to Fig.~\ref{fig:f3_1e3} but for an IMBH with $m_I=10^2\, M_\odot$ and $P_b^I=3\,{\rm yr}$. \label{fig:f2_1e3} }
	\end{center}
\end{figure}

It is also interesting whether timing observation of the pulsar-SMBH-IMBH system can measure the 1\,PN three-body interaction term as a unique test of GR. By adding the additional parameter $f_3$ in the timing model, we can obtain its measurement precision via a similar procedure as before. The estimation results for all the examples are shown in Fig.~\ref{fig:f3_1e3} and Fig.~\ref{fig:f2_1e3}. Our results suggest that the measurement precision of $f_3$ is rather low for the allowed mass range of the IMBH.

\section{Detectability of the IMBH}\label{sec:detect}

The discussions in the previous section are based on the assumption of an ideal three-body system consisting only of the pulsar, SMBH, and IMBH. However, for real observations, the complex environment in the GC may spoil the measurement of the IMBH parameter. Specifically, the high measurement precision of the IMBH mass expected in the last section suggests that any star that has an orbital period similar to the pulsar will, in principle, affect the timing data.  Therefore, the long predicted stellar cusp around Sgr~A* may cause unexpected problems in the timing observation of the pulsar-SMBH system, as we explicitly showed in Ref.~\cite{Hu:2026aez}.

Nevertheless, one may still detect the IMBH if it provides an outstanding signal that is stronger than the background noise in the timing residuals. Similar to Sec.~\ref {subsec:post-fit}, here we consider the amplitude of the post-fit timing residual caused by the IMBH. To estimate the detectability of an IMBH with a given orbital period (or equivalently, the orbital semi-major axis), we simulate 1000 systems with IMBHs having random orbital parameters but with fixed mass $m_I$ and orbital period $P_b^I$. We calculate the post-fit timing residuals for each system and count the number of systems with a maximum post-fit timing residual larger than a given threshold. By changing the mass $m_I$, we can find a value, e.g., $m_I^{95}$, which is the lowest mass that for an IMBH with this mass and the given orbital period, one can see a signal in the post-fit timing residuals with its peak amplitude larger than the given threshold in $95\%$ simulations. 

To generate the random systems used for the above calculation, we draw samples of the IMBH orbital parameters as follows. The orbital orientation controlled by $\Omega^I$, $\omega^I$, and $i^I$ is uniformly distributed in the $4\pi$ solid angle. For the initial orbital phase $f_0^I$, we draw samples so that the number of systems in $\big[f_0^I,f_0^I+{\rm d}f_0^I \big]$ is proportional to the time that the IMBH passes this part of the orbit; or equivalently, one can directly draw uniform samples for $T_0$, the time epoch of periastron passage. Finally, for the eccentricity $e_I$, we assume a uniform distribution between $0.1$ and $0.9$.

The threshold for the amplitude of the post-fit timing residuals is related to the background noise caused by other perturbations. As studied in Ref.~\cite{Hu:2026aez}, a stellar mass BH cusp expected by the stellar dynamics and satisfying the current observational constraints~\cite{GRAVITY:2024tth} can lead to timing residuals as large as $10^1$--$10^2\,{\rm s}$ depending on the total mass of the BH cusp. Therefore, we consider two thresholds, $10\,{\rm s}$ and $100\,{\rm s}$ as illustrations. 

\begin{figure}[H]
	\begin{center}
		\includegraphics[width=8.5cm]{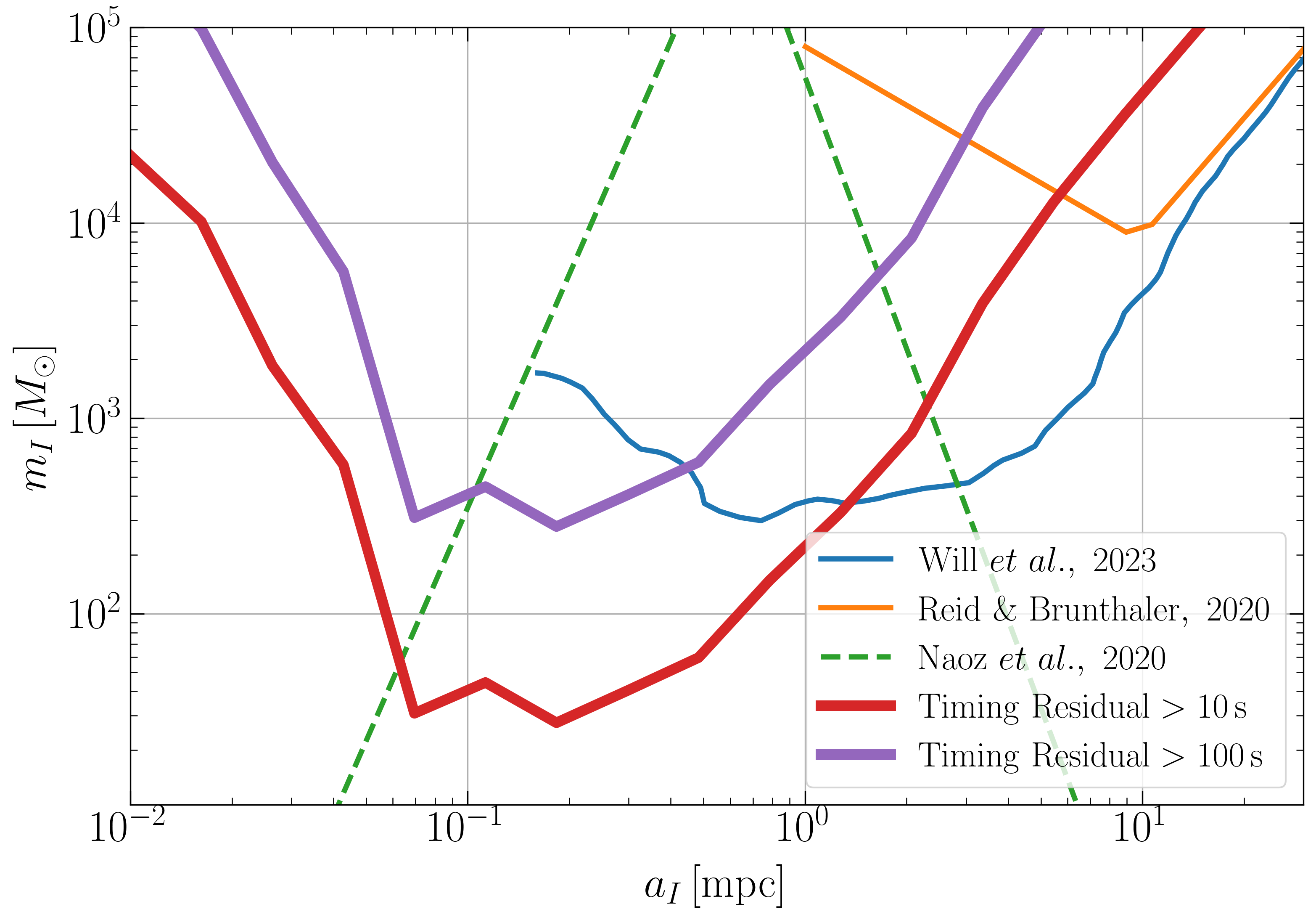}
	\caption{ Detectability of an IMBH estimated in this work and current constraints on the existence of an IMBH from various studies. Parameter space above the curves is excluded. \label{fig:detect} }
	\end{center}
\end{figure}

In Fig.~\ref{fig:detect}, we show the $m_I^{95}$ estimated with the above procedure as 
functions of the IMBH orbital semi-major axis. For comparison, we also show the current 
constraints on the existence of an IMBH near the Sgr~A*. Will \textit{et al.}~\cite{Will:2023nlt} 
constrained the companion of Sgr~A* based on the observations of the S0-2 star; 
a more complex constraint obtained by Straub \textit{et al.}~\cite{GRAVITY:2023met} 
is not shown in the figure. The constraints by Reid and Brunthaler~\cite{Reid:2020} are 
based on the proper motion measurement of the Sgr~A*. The dashed line in the figure is 
related to the constraints estimated by Naoz \textit{et al.}~\cite{Naoz:2019sjx} based on 
the stability of the S-star cluster. From the figure, one can see that timing a pulsar orbiting around Sgr~A* can provide constraints on the existence of an IMBH in a unique parameter space depending on the pulsar's orbital period. Even with large environmental perturbations, a proper pulsar orbiting around Sgr~A* can still provide valuable information for finding or constraining the possible companion of Sgr~A*.

\section{Discussion}\label{sec:discussion}

In this paper, we study the timing observation of a pulsar-SMBH-IMBH system that was speculated in our GC. We construct a timing model for this three-body system based on the canonical equations of motion derived from the 1\,PN Hamiltonian with the leading-order spin-orbit interaction. We include various leading-order timing delays in this system, which is enough for our purpose of parameter estimation, but should be extended for a realistic timing model used in  observations~\cite{Hu:2026zcb}. Our timing model is designed so that it might also be applicable to the triple-pulsar system by moderate modifications.

As for theoretical interests, we first study an ideal pulsar-SMBH-IMBH system in the sense that there are no other mass perturbations. We present the various pre-fit time delays and post-fit timing residuals caused by the IMBH, which might be helpful for determining the existence of an IMBH from timing residuals. Shapiro delay can be a distinct signature if the three-body system is in some special configuration. Otherwise, one might search for signals modulated by double periods in the post-fit timing residual if the observation time span is long enough. 

We also perform parameter estimation based on the Fisher matrix method. For an ideal system, the IMBH could cause a large post-fit timing residual compared to the expected timing precision of GC pulsars. Therefore, one would expect that the measurement of the IMBH parameters can reach unprecedented precision, which is consistent with our numerical results. However, our calculation in fact suggests that any significant stellar mass objects orbiting around Sgr~A* with a similar orbital period to that of the pulsar should be taken into account in real analysis. The stellar cusp around the SMBH predicted by the stellar dynamics may cause serious problems in future timing observations of pulsars orbiting around Sgr~A*, as we discussed thoroughly in another study~\cite{Hu:2026aez}. Ways to overcome the difficulties were also presented there.

Similarly, various environmental effects in the GC need to be justified. For example, in Ref.~\cite{Hu:2023ubk}, we 
studied the perturbations caused by the dark matter. Gas around Sgr~A* may also introduce addition effects in the pulsar's 
orbital motion and even affect the pulsar's emission process. We leave these aspects for future studies.

Nevertheless, one may still expect to detect or constrain the existence of an IMBH if it can provide an outstanding signal compared to the background noise. By simulating systems with different orbital configurations, we estimate the minimum IMBH mass that can cause post-fit timing residuals larger than a given threshold for most cases. With 10\,s and 100\,s as example thresholds~\cite{Hu:2026aez}, our simulation suggests that, even with a strong environmental perturbation, a proper pulsar orbiting around Sgr~A* can still provide valuable constraints on the existence of an IMBH and vastly complement existing observational constraints.

\Acknowledgements{We thank Norbert Wex for discussions. This work
was supported by the National Natural Science Foundation
of China (124B2056, 12573042), the National SKA Program of China (2020SKA0120300), the Beijing Natural Science Foundation (1242018), the Max Planck Partner Group
Program funded by the Max Planck Society, and the High-performance Computing Platform of Peking University.}

\InterestConflict{The authors declare that they have no conflict of interest.}


\bibliographystyle{scpma}
\bibliography{refs.bib}

\begin{thebibliography}{10}
\providecommand{\url}[1]{\texttt{#1}}
\providecommand{\urlprefix}{URL }
\providecommand{\doi}[1]{doi:~\href{http://doi.org/#1}{\nolinkurl{#1}}}
\providecommand{\arXiv}[1]{\href{https://arxiv.org/abs/#1}{\nolinkurl{https://arxiv.org/abs/#1}}}
\providecommand{\eprint}[1]{\href{http://arxiv.org/abs/#1}{\nolinkurl{#1}}}

\bibitem{Ghez:2003rt}
A.~M. Ghez et~al., Astrophys. J. Lett. \textbf{586}, L127 (2003).

\bibitem{Alexander:2008tq}
T.~Alexander and C.~Hopman, Astrophys. J. \textbf{697}, 1861 (2009).

\bibitem{Hansen:2003yb}
B.~Hansen and M.~Milosavljevic, Astrophys. J. Lett. \textbf{593}, L77 (2003).

\bibitem{Yu:2007jz}
Q.~Yu, Y.~Lu, and D.~N.~C. Lin, Astrophys. J. \textbf{666}, 919 (2007).

\bibitem{Zheng:2026}
X.~{Zheng}, L.~{Wang}, D.~N.~C. {Lin}, A.~{Burkert}, and S.~{Mao}, Astrophys. J. \textbf{1005}, 97 (2026), arXiv: \eprint{2606.08971}.

\bibitem{Chen:2014dya}
X.~Chen and P.~Amaro-Seoane, Astrophys. J. Lett. \textbf{786}, L14 (2014).

\bibitem{Hopkins:2005fb}
P.~F. Hopkins, L.~Hernquist, T.~J. Cox, T.~Di~Matteo, B.~Robertson, and V.~Springel, Astrophys. J. Suppl. \textbf{163}, 1 (2006).

\bibitem{Rashkov:2013uua}
V.~Rashkov and P.~Madau, Astrophys. J. \textbf{780}, 187 (2014).

\bibitem{Naoz:2019sjx}
S.~Naoz, C.~M. Will, E.~Ramirez-Ruiz, A.~Hees, A.~M. Ghez, and T.~Do, Astrophys. J. Lett. \textbf{888}, L8 (2020).

\bibitem{Zhang:2023ekp}
E.~Zhang, S.~Naoz, and C.~M. Will, Astrophys. J. \textbf{952}, 103 (2023).

\bibitem{Will:2023nlt}
C.~M. Will, S.~Naoz, A.~Hees, A.~Tucker, E.~Zhang, T.~Do, and A.~Ghez, Astrophys. J. \textbf{959}, 58 (2023).

\bibitem{GRAVITY:2023met}
O.~Straub et~al. (GRAVITY), Astron. Astrophys. \textbf{672}, A63 (2023), [Erratum: Astron.Astrophys. 677, C2 (2023)].

\bibitem{Yu:2003hj}
Q.~Yu and S.~Tremaine, Astrophys. J. \textbf{599}, 1129 (2003).

\bibitem{Reid:2020}
M.~J. Reid and A.~Brunthaler, Astrophys. J. \textbf{892}, 39 (2020).

\bibitem{Strokov:2023kmo}
V.~Strokov, G.~Fragione, and E.~Berti, Mon. Not. Roy. Astron. Soc. \textbf{524}, 2033 (2023).

\bibitem{Guo:2024tlg}
X.~Guo, Q.~Yu, and Y.~Lu, Astrophys. J. \textbf{978}, 104 (2025), arXiv: \eprint{2411.14150}.

\bibitem{Wex:1998wt}
N.~Wex and S.~Kopeikin, Astrophys. J. \textbf{514}, 388 (1999), arXiv: \eprint{astro-ph/9811052}.

\bibitem{Kramer:2004hd}
M.~Kramer, D.~C. Backer, J.~M. Cordes, T.~J.~W. Lazio, B.~W. Stappers, and S.~Johnston, New Astron. Rev. \textbf{48}, 993 (2004), arXiv: \eprint{astro-ph/0409379}.

\bibitem{Liu:2011ae}
K.~Liu, N.~Wex, M.~Kramer, J.~M. Cordes, and T.~J.~W. Lazio, Astrophys. J. \textbf{747}, 1 (2012).

\bibitem{Psaltis:2015uza}
D.~Psaltis, N.~Wex, and M.~Kramer, Astrophys. J. \textbf{818}, 121 (2016).

\bibitem{Hu:2023ubk}
Z.~Hu, L.~Shao, and F.~Zhang, Phys. Rev. D \textbf{108}, 123034 (2023).

\bibitem{Dong:2022zvh}
Y.~Dong, L.~Shao, Z.~Hu, X.~Miao, and Z.~Wang, JCAP \textbf{11}, 051 (2022), arXiv: \eprint{2210.16130}.

\bibitem{Hu:2023vsg}
Z.~Hu, L.~Shao, R.~Xu, D.~Liang, and Z.-F. Mai, JCAP \textbf{04}, 087 (2024), arXiv: \eprint{2312.02486}.

\bibitem{DellaMonica:2023ydm}
R.~Della~Monica, I.~de~Martino, and M.~de~Laurentis, Mon. Not. Roy. Astron. Soc. \textbf{524}, 3782 (2023), arXiv: \eprint{2305.18178}.

\bibitem{Hu:2024blq}
Z.~Hu and L.~Shao, Phys. Rev. Lett. \textbf{133}, 231402 (2024), arXiv: \eprint{2408.00245}.

\bibitem{Shao:2025vmb}
L.~Shao and Z.~Hu, J. Phys. Conf. Ser. \textbf{3177}, 012043 (2026), arXiv: \eprint{2508.09931}.

\bibitem{DellaMonica:2025ent}
R.~Della~Monica and I.~de~Martino, Phys. Rev. D \textbf{114}, 024048 (2026), arXiv: \eprint{2501.03912}.

\bibitem{Yu:2025apk}
J.-C. Yu, Y.~Cao, Z.~Hu, and L.~Shao  (2025), arXiv: \eprint{2510.22573}.

\bibitem{Hu:2026zcb}
Z.~Hu, Z.~Wang, and L.~Shao, Astrophys. J. Suppl. \textbf{286}, 41 (2026), arXiv: \eprint{2602.19546}.

\bibitem{Hu:2026aez}
Z.~Hu and L.~Shao  (2026), arXiv: \eprint{2606.04762}.

\bibitem{Pfahl:2003tf}
E.~Pfahl and A.~Loeb, Astrophys. J. \textbf{615}, 253 (2004).

\bibitem{Zhang:2014kva}
F.~Zhang, Y.~Lu, and Q.~Yu, Astrophys. J. \textbf{784}, 106 (2014).

\bibitem{Schoedel2020}
R.~Sch{\"o}del et~al., Astron.Astrophys. \textbf{641}, A102 (2020).

\bibitem{SKAOPulsarScienceWorkingGroup:2025syv}
F.~Abbate et~al. (SKAO Pulsar Science Working Group), Open J. Astrophys. \textbf{8}, 54252 (2025), arXiv: \eprint{2512.16155}.

\bibitem{Johnston:2006fx}
S.~Johnston, M.~Kramer, D.~R. Lorimer, A.~G. Lyne, M.~McLaughlin, B.~Klein, and R.~N. Manchester, Mon. Not. Roy. Astron. Soc. \textbf{373}, L6 (2006).

\bibitem{Deneva:2009mx}
J.~S. Deneva, J.~M. Cordes, and T.~J.~W. Lazio, Astrophys. J. Lett. \textbf{702}, L177 (2009).

\bibitem{Eatough:2013nva}
R.~P. Eatough et~al., Nature \textbf{501}, 391 (2013).

\bibitem{Rea:2013pqa}
N.~Rea et~al., Astrophys. J. Lett. \textbf{775}, L34 (2013).

\bibitem{Wongphechauxsorn:2023qcy}
J.~Wongphechauxsorn et~al., Mon. Not. Roy. Astron. Soc. \textbf{527}, 3208 (2023).

\bibitem{Yao2017}
J.~M. Yao, R.~N. Manchester, and N.~Wang, Astrophys. J. \textbf{835}, 29 (2017).

\bibitem{Ocker:2026mta}
S.~K. Ocker and J.~M. Cordes, Astrophys. J. \textbf{1002}, 3 (2026).

\bibitem{Liu:2021ziv}
K.~Liu et~al., Astrophys. J. \textbf{914}, 30 (2021), arXiv: \eprint{2104.08986}.

\bibitem{EventHorizonTelescope:2023atv}
P.~Torne et~al. (Event Horizon Telescope), Astrophys. J. \textbf{959}, 14 (2023), arXiv: \eprint{2308.15381}.

\bibitem{Kramer:2021jcw}
M.~Kramer et~al., Phys. Rev. X \textbf{11}, 041050 (2021).

\bibitem{Hu:2023vsq}
Z.~{Hu}, X.~{Miao}, and L.~{Shao}, in \emph{Recent Progress on Gravity Tests. Challenges and Future Perspectives}, (edited by C.~{Bambi} and A.~{C{\'a}rdenas-Avenda{\~n}o}), 61--99 (2024), arXiv: \eprint{2303.17185}.

\bibitem{Damour:1985}
T.~Damour and N.~Deruelle, Ann.~Inst.~Henri Poincar{\'e} Phys.~Th{\'e}or. \textbf{43}, 107 (1985).

\bibitem{Shao:2016ubu}
L.~Shao, Phys. Rev. D \textbf{93}, 084023 (2016), arXiv: \eprint{1602.05725}.

\bibitem{Voisin:2020lqi}
G.~Voisin, I.~Cognard, P.~C.~C. Freire, N.~Wex, L.~Guillemot, G.~Desvignes, M.~Kramer, and G.~Theureau, Astron. Astrophys. \textbf{638}, A24 (2020).

\bibitem{Heinze:2026rho}
F.~M. Heinze, G.~Sch{\"a}fer, and B.~Br{\"u}gmann, Phys. Rev. D \textbf{113}, 104066 (2026).

\bibitem{Einstein:1938yz}
A.~Einstein, L.~Infeld, and B.~Hoffmann, Annals Math. \textbf{39}, 65 (1938).

\bibitem{Will:2013cza}
C.~M. Will, Phys. Rev. D \textbf{89}, 044043 (2014), [Erratum: Phys.Rev.D 91, 029902 (2015)].

\bibitem{Damour:1986}
T.~Damour and N.~Deruelle, Ann.~Inst.~Henri Poincar{\'e} Phys.~Th{\'e}or. \textbf{44}, 263 (1986).

\bibitem{Lorimer:2004handbook}
D.~R. Lorimer and M.~Kramer, \emph{Handbook of Pulsar Astronomy} (2005).

\bibitem{Pryce:1948pf}
M.~H.~L. Pryce, Proc. Roy. Soc. Lond. A \textbf{195}, 62 (1948).

\bibitem{Newton:1949cq}
T.~D. Newton and E.~P. Wigner, Rev. Mod. Phys. \textbf{21}, 400 (1949).

\bibitem{Wex:1995pjg}
N.~Wex, Class. Quant. Grav. \textbf{12}, 983 (1995).

\bibitem{poisson_will_2014}
E.~Poisson and C.~M. Will, \emph{Gravity: Newtonian, Post-Newtonian, Relativistic} (2014).

\bibitem{Damour:1988mr}
T.~Damour and G.~Sch{\"a}fer, Nuovo Cim. B \textbf{101}, 127 (1988).

\bibitem{Blandford:1976ApJ}
R.~Blandford and S.~A. Teukolsky, Astrophys. J. \textbf{205}, 580 (1976).

\bibitem{Shapiro:1964uw}
I.~I. Shapiro, Phys. Rev. Lett. \textbf{13}, 789 (1964).

\bibitem{Will:2018mcj}
C.~M. Will, Phys. Rev. Lett. \textbf{120}, 191101 (2018).

\bibitem{Edwards:2006zg}
R.~T. Edwards, G.~B. Hobbs, and R.~N. Manchester, Mon. Not. Roy. Astron. Soc. \textbf{372}, 1549 (2006).

\bibitem{Zhang:2017qbb}
F.~Zhang and P.~Saha, Astrophys. J. \textbf{849}, 33 (2017).

\bibitem{Gillessen:2017jxc}
S.~Gillessen et~al., Astrophys. J. \textbf{837}, 30 (2017).

\bibitem{Peebles:1972}
P.~J.~E. Peebles, Astrophys. J. \textbf{178}, 371 (1972).

\bibitem{GRAVITY:2024tth}
K.~Abd El~Dayem et~al. (GRAVITY), Astron. Astrophys. \textbf{692}, A242 (2024).

\end{thebibliography}


\end{multicols}
\end{document}